\documentclass[reprint,aps,
pre,
floatfix,
notitlepage,
superscriptaddress
]{revtex4-2}
\usepackage{amssymb,amsmath,bm,bbm}
\usepackage{graphicx}
\usepackage{float}
\usepackage{multirow}
\usepackage{xcolor}
\usepackage[colorlinks=true,allcolors=blue]{hyperref}
\usepackage{comment}
\usepackage{mathtools}

\usepackage{mathrsfs}

\usepackage{ragged2e}
\justifying

\usepackage{soul}
\definecolor{linkcolor}{HTML}{004191}
\usepackage{physics}

\newcommand{\RM}[1]{\mathrm{#1}}
\newcommand{\p}{\partial}

\begin{document}

\title{Designing topological edge currents in chiral active matter}

\author{Yuta Kuroda}
\email{yuta.kuroda@itp4.uni-stuttgart.de}
\affiliation{Institute for Theoretical Physics IV, University of Stuttgart, Heisenbergstr.~3, 70569 Stuttgart, Germany}

\author{Ellen Meyberg}
\affiliation{Institute for Theoretical Physics IV, University of Stuttgart, Heisenbergstr.~3, 70569 Stuttgart, Germany}

\author{Gaurav Gardi}
\affiliation{Physical Intelligence Department, Max Planck Institute for Intelligent Systems, 70569 Stuttgart, Germany}

\author{Thomas Speck}
\email{thomas.speck@itp4.uni-stuttgart.de}
\affiliation{Institute for Theoretical Physics IV, University of Stuttgart, Heisenbergstr.~3, 70569 Stuttgart, Germany}

\author{Saeed Osat}
\email{saeedosat13@gmail.com}
\affiliation{Institute for Theoretical Physics IV, University of Stuttgart, Heisenbergstr.~3, 70569 Stuttgart, Germany}

\begin{abstract}
Achieving robust functionality in active matter driven away from thermal equilibrium is a current theoretical and experimental challenge.
Several recent studies have reported edge currents--persistent transport along walls and density inhomogeneities--in chiral active matter.
Yet, the microscopic rules that render these edge currents robust with respect to the confinement geometry and defects remain elusive.
Here, we introduce a simple particle model of two-dimensional chiral active swimmers that undergo chirality switching and demonstrate that the model exhibits robust edge currents, i.e., when a single particle is confined, edge currents arise regardless of the confinement geometry or the presence of defects.
We also investigate the collective behavior of interacting particles in bulk and find that chirality switching induces phase separation accompanied by edge currents along interfaces.
This phase separation is distinct from motility-induced phase separation and is qualitatively explained by an effective hydrodynamic theory derived via bottom-up coarse-graining.
Furthermore, by analyzing the topological properties of the linearized hydrodynamic equations, we show that the edge currents in our system are genuine topological edge modes.
Notably, phase separation induced by chirality switching can be regarded as the coexistence of two topologically distinct domains.
Our results provide guidelines for designing robust edge currents in active matter systems.
\end{abstract}

\maketitle

\section{Introduction}
Biological systems operate far from equilibrium, where the established framework of equilibrium thermodynamics breaks down~\cite{Qian2007}.
The physics of active matter has expanded our understanding of such systems and their synthetic counterparts~\cite{Marchetti2013, Ramaswamy2010, Bechinger2016RevModPhys, LesHouches2018}.
Examples include, but are not limited to, flocks~\cite{Vicsek1995, Cavagna2014AnnuRev}, bacterial suspensions~\cite{Dombrowski2004PRL}, active gels~\cite{Sanchez2012}, and self-propelled colloidal particles~\cite{Volpe2011SoftMat, Buttinoni2013PRL, Nishiguchi2015PRE}. 
Another key feature of biological systems is their robust functionality in crowded and noisy environments~\cite{Kitano2004}.
Understanding the origin of this robustness is essential not only for explaining biological organization, but also for endowing engineered synthetic systems with similar life-like capabilities.

Among the various types of active matter systems, chiral active matter has attracted much attention~\cite{Lowen2016EPJ, liebchen2022chiral}. 
In such chiral systems, each constituent undergoes motion that breaks mirror symmetry.
In two dimensions, one typical manifestation of chiral motion is self-propulsion biased toward the left or right, leading to circular trajectories.
Experimental realizations of such chiral self-propelled motion include asymmetric colloidal objects~\cite{Kummel2013PRL, Zhang:2020ux}, Janus particles with asymmetric coatings~\cite{Mano2017}, sperm cells~\cite{Ingmar2005Science}, malaria parasites~\cite{Patra:2022vd}, bacteria swimming near walls or surfaces~\cite{DiLuzio2005, Leonardo2011}, and light-driven walkers~\cite{Siebers2023}.
Collections of chiral active particles are known to exhibit a wealth of intriguing phenomena that are absent in their achiral counterparts.
Examples include vortex formation~\cite{Liebchen2017, Kruk2020PRE, Ventejou2021PRL, Liao2021SoftMatter}, absorbing phase transitions~\cite{Lei2019}, dynamical hyperuniformity~\cite{Lei2019, Huang2021, Zhang2022, Kuroda2023JStatMech, Kuroda2025PRE}, reentrant behavior of glassy slow dynamics~\cite{Debets2023PRL}, anomalous two-dimensional crystallization~\cite{kuroda2025PRR, Jeong2025PRR}, and self-reverting vorticity of clusters~\cite{caprini2024self}.

In this context of chiral active matter, ``odd'' transport phenomena have also become a major research focus~\cite{Fruchart2023AnnuRev}.
One of the most notable consequences of such odd transport is the emergence of edge currents in systems with interfaces, such as confined systems~\cite{Benjamin2015PNAS, Negi2023PRR, Alsallom2026} or phase-separated states~\cite{Caporusso2024PRL, Adorjani2024EPJ, Siebers2024PNAS, Zhou2025PNASNex, Caprini2025JCP, Metzger2026, wang2026edge}.
Experimentally, edge currents have been reported, for example, in granular matter~\cite{Tsai2005PRL, Caprini2025arXiv, Goerlich2026}, magnetic colloids~\cite{Tierno2007PRL, Soni2019NatPhys, Massana-Cid2021PRR, Katuri2024CommPhys}, Quincke rollers~\cite{Bricard2015NatComm}, synthetic active spinners~\cite{Workamp2018SoftMat, Liu2020PNAS, Yang2020PRE}, swarming bacteria~\cite{beppu2021edge, Li2024PRX}, and living cells~\cite{Yashunsky2022PRX}.
Of particular interest is the robustness of such edge currents, and several studies have suggested that the currents persist despite changes in interface geometry~\cite{Yang2020PRE, Li2024PRX}.

Meanwhile, topological band theory provides a powerful theoretical framework for understanding robust transport properties in condensed matter systems.
Although this framework was originally developed in the context of quantum condensed matter systems such as quantum Hall systems~\cite{Klitzing1980PRL, Thouless1982PRL, Kohmoto1985} and topological insulators~\cite{Hasan2010RMP, FRUCHART2013779}, it is now known to be relevant to a wide range of physical systems, including photonics~\cite{Lu2014, Ozawa2019}, metamaterials~\cite{NashPNAS2015, Ni2023, Roman2016PNAS}, stochastic systems~\cite{Murugan2017NatComm, Knebel2020PRL, Sawada2024PRL}, colloidal systems~\cite{Loehr2018CommPhys}, and active matter~\cite{Dasbiswas2018PNAS, Souslov2017, Sone2019PRL, Souslov2019PRL, Green2020, Sone2020NatComm, Yamauchi2020, Yang2021PRL, Palacios2021NatComm, Shankar2022NRP, Sone_2026, Uchida2026, wojcik2026chiral, Osat2026}.
In particular, Souslov~{\it et al.}~\cite{Souslov2019PRL} suggested the existence of topologically protected edge modes in chiral active fluids by exploiting a mathematical analogy between topological band theory for Hermitian systems and a continuum description of compressible fluids with odd viscosity.
From a microscopic point of view, topological edge modes have been predicted in lattice models of a single chiral active particle~\cite{wojcik2026chiral, Osat2026}.
Yet, the link between microscopic interactions and the emergent topological modes predicted by large-scale hydrodynamic theories remains largely unexplored.

This gap motivates us to construct a simple particle model that exhibits robust edge currents, allowing us to examine whether, and if so how, these currents are connected to topological edge modes through coarse-graining.
Inspired by recent observations suggesting that chirality switching may be key to realizing robust edge currents~\cite{Osat2026}, we consider chiral active Brownian particles augmented with a feedback term that can change the direction of circular motion.
Such chirality switching has been observed in several experimental setups.
For example, {\it E. coli} is known to change its chirality depending on boundary conditions~\cite{Lemelle2010, Lemelle2013SoftMat}; the chirality of sea urchin sperm cells is controlled by calcium ions~\cite{Ishijima1993}; and {\it Chlamydomonas} has been reported to switch its chirality depending on light intensity~\cite{Wang2026PRL}.
We first show that, even at the single-particle level, the model exhibits edge currents that are robust against changes in confinement geometry and the presence of defects.
We then numerically examine the collective behavior of interacting particles in bulk and find that the system spontaneously separates into dense and dilute regions, akin to liquid-gas phase separation.
Remarkably, edge currents emerge along the interfaces of droplets.
Moreover, we show that phase separation in our system does not follow the mechanism of motility-induced phase separation~\cite{Cates2015AnnuRev}, but is instead induced by chirality switching.

In both the single- and many-body cases, we account for the emergence of edge currents in two steps.
(i)~We develop an effective hydrodynamic description via a bottom-up approach.
The derived equations contain odd transport terms, which qualitatively explain the existence of edge currents.
However, this description does not yet explain the robustness of the edge currents.
(ii)~To explain this robustness, we analyze the topological properties of the system.
By linearizing the hydrodynamic equations, one can map the system onto a three-level system described by an effective non-Hermitian Hamiltonian.
We then calculate the bulk topological invariants and find that, as long as the bulk gap remains open, the topology is determined solely by the sign of the effective chirality.
For a single particle under confinement, the robust edge currents can be understood as topological edge modes that emerge at the boundary between two topologically distinct domains.
In the phase-separated states observed in the many-body case, we find that the dense and dilute phases are topologically distinct.
We therefore identify the interfacial currents in our model as topological edge modes associated with the spontaneous formation of a topological domain wall between the two phases.
We also calculate the band structure of the effective Hamiltonian under open boundary conditions and show the existence of gapless edge modes, supporting the bulk-boundary correspondence in our model.

This paper is organized as follows.
We introduce our particle-based model of active swimmers with chirality switching in Sec.~\ref{sec:model}.
In Sec.~\ref{sec:single-particle}, we study the single-particle behavior in the presence of walls and defects and show that the model exhibits robust edge currents.
Subsequently, the collective behavior of interacting particles is investigated in Sec.~\ref{sec:collective_behavior}.
We show that chirality switching induces phase separation accompanied by edge currents localized near interfaces.
We then develop a hydrodynamic theory for our particle model and qualitatively explain the phase separation with edge currents observed in the numerical simulations.
In Sec.~\ref{sec:topo}, we discuss the edge currents in our system from the perspective of topology.
Finally, Sec.~\ref{sec:summary} concludes the paper. 

\section{Model}
\label{sec:model}

\begin{figure}[b!]
\centering
\includegraphics[width=1.0\linewidth]{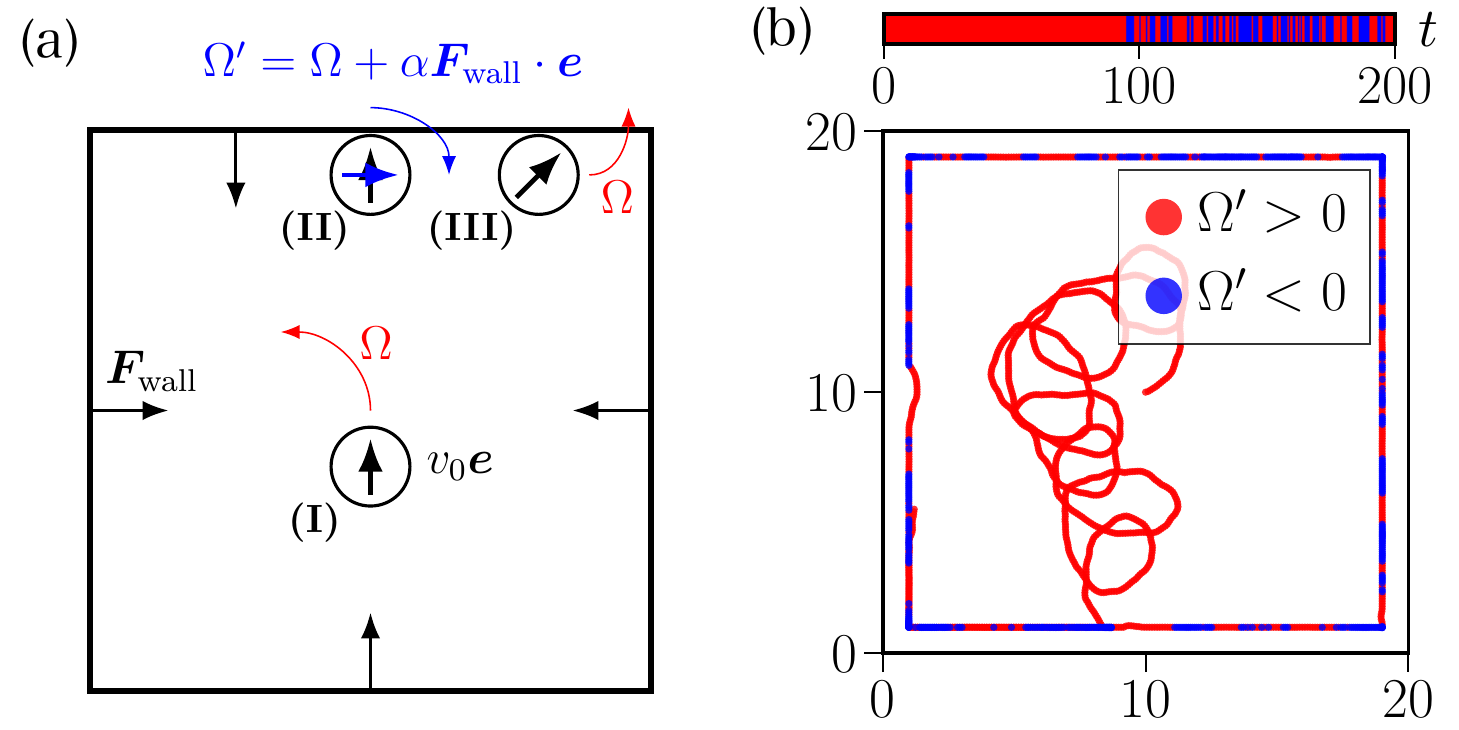}
\caption{\label{fig:model}A chiral active particle with linear speed $v_0$ and angular speed $\Omega$. (a)~Illustration of chirality switching induced by particle-wall interactions. 
See main text for discussion of processes (I-III).
(b)~Illustration of the edge current induced by chirality switching.
The particle is confined in a square box and initially placed at the center of the box.
The particle trajectory is colored according to the sign of the effective torque $\Omega'(t)$.
The top bar shows the corresponding time evolution of $\RM{sgn}[\Omega'(t)]$.
The values $\RM{sgn}[\Omega'(t)] = +1$ and $-1$ are shown in red and blue, respectively.
The parameters are set to $\Omega = 0.5$, $\alpha =2$, and $D_\RM{r}=0.1$.
}
\end{figure}
\begin{figure*}
\centering
\includegraphics[width=0.9\linewidth]{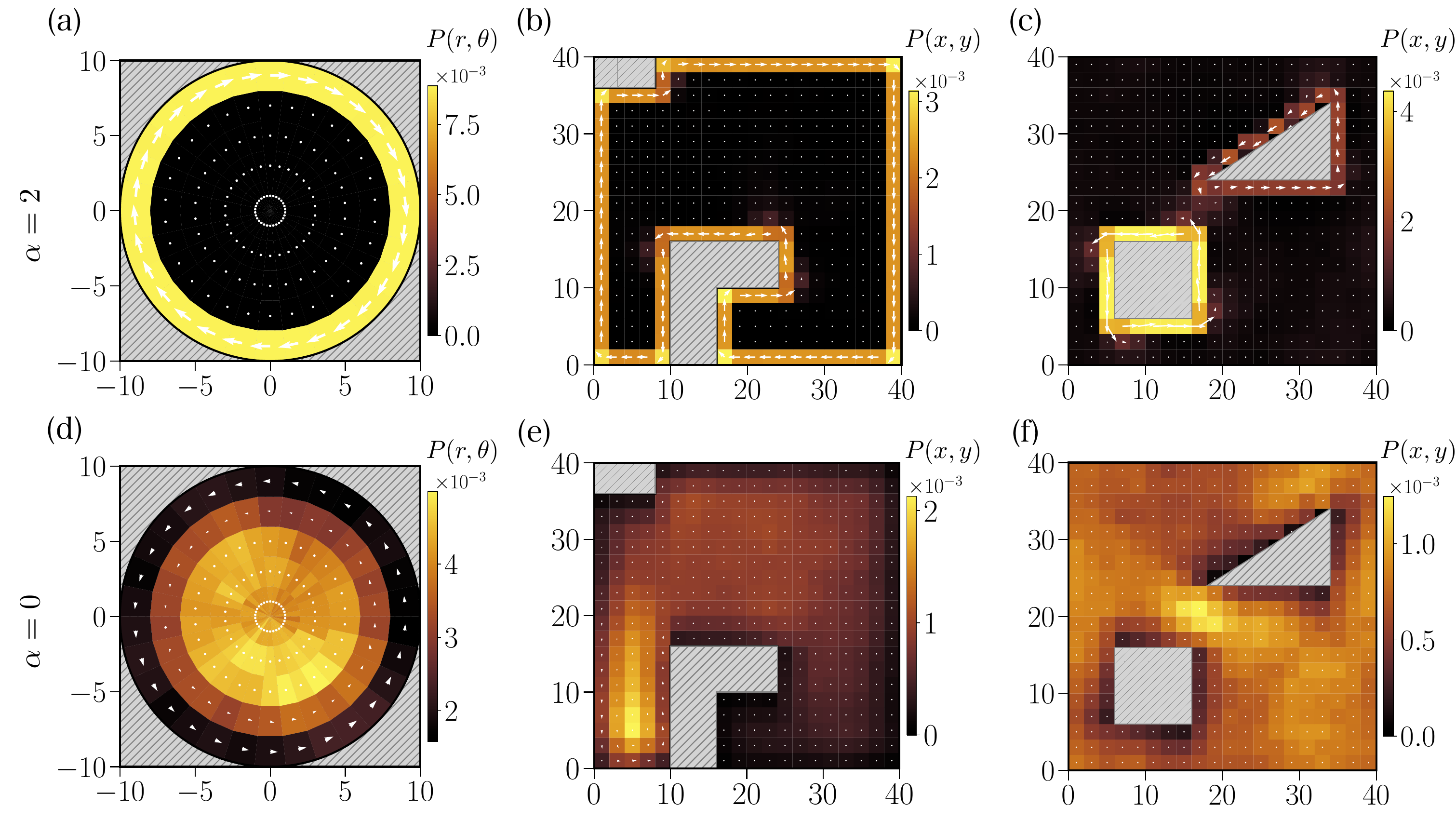}
\caption{\label{fig:histogram}
Single particle edge currents in three different wall and defect geometries.
The upper row, panels (a)-(c), and the lower row, panels (d)-(f), correspond to $\alpha=2$ and $\alpha=0$, respectively.
From left to right, the panels show the spatial distributions of a particle in circular confinement, confinement with a complex boundary, and a system with two defects under periodic boundary conditions.
In each panel, the white arrows indicate the density current.
The spatial distributions and density currents are calculated by discretizing the space into small cells.
All other parameters are fixed at $\Omega =0.5$ and $D_\RM{r}=0.1$.
}
\end{figure*}
We study a model of two-dimensional chiral self-propelled particles that exhibits chirality switching due to feedback from interactions (Fig.~\ref{fig:model}).
The translational motion of the $j$-th particle is assumed to be an overdamped equation
\begin{equation}
\frac{\dd \bm r_j(t)}{\dd t} = \mu \bm F_j(t) + v_0\bm e(\phi_j(t)) \label{eom1}, 
\end{equation}
for the position $\bm r_j(t)$, where $\mu$ is the mobility and
$\bm F_j(t)$ is the total force acting on the $j$-th particle. 
The last term, $v_0\bm e(\phi)=v_0(\cos\phi,\sin\phi)$, represents self-propulsion with speed $v_0$.
The orientational angle obeys the stochastic equation
\begin{equation}
\frac{\dd \phi_j(t)}{\dd t} = \Omega + T_j(t)+ \sqrt{2D_\RM{r} }\eta_j(t), \label{eom2}  
\end{equation}
where $\eta_j(t)$ is a Gaussian white noise with zero mean and unit variance, and $D_\RM{r}$ represents the rotational diffusion constant. 
$\Omega$ is a constant chiral torque that leads to a circular motion.
The second term, $T_j(t)$, indicates an additional torque.
To achieve switching of the effective chirality when a particle approaches walls or other particles, we employ
\begin{equation}
T_j(t) = \alpha \bm F_j(t) \cdot \bm e(\phi_j(t)), \label{eq:def-torque}
\end{equation}
where the parameter $\alpha$ controls the strength of the torque. 
The role of this term is illustrated in Fig.~\ref{fig:model}(a), where we consider a single particle under confinement.
When the particle is away from the wall, $T_j(t) = 0$, and the particle performs a counterclockwise circular motion for $\Omega>0$~[(I) in Fig.~\ref{fig:model}(a)].
In this case, the particle motion is identical to that of a well-studied chiral active Brownian particle, whose long-time behavior is diffusive and characterized by the diffusion coefficient $\mathcal D = v_0^2 D_\RM{r}/[2\qty(D_\RM{r}^2 + \Omega^2)]$~\cite{vanTeeffelen2008PRE}.
In the absence of noise, $D_\RM{r}=0$, the particle trajectory is a perfect circle of radius $R=v_0/\Omega$.
When the particle collides with the wall with its orientation pointing toward the wall, it experiences a negative torque $T_j <0$, which can switch the direction of rotation~[(II) in Fig.~\ref{fig:model}(a)] if $\alpha$ is sufficiently large.
If the particle then moves slightly away from the wall, it starts to rotate under the torque $\Omega$ and collides with the wall again~[(III) in Fig.~\ref{fig:model}(a)].
As a result, processes (II) and (III) repeat near the wall, leading to an edge current [Fig.~\ref{fig:model}(b)].

\begin{figure}[b!]
\centering
\includegraphics[width=8.5cm]{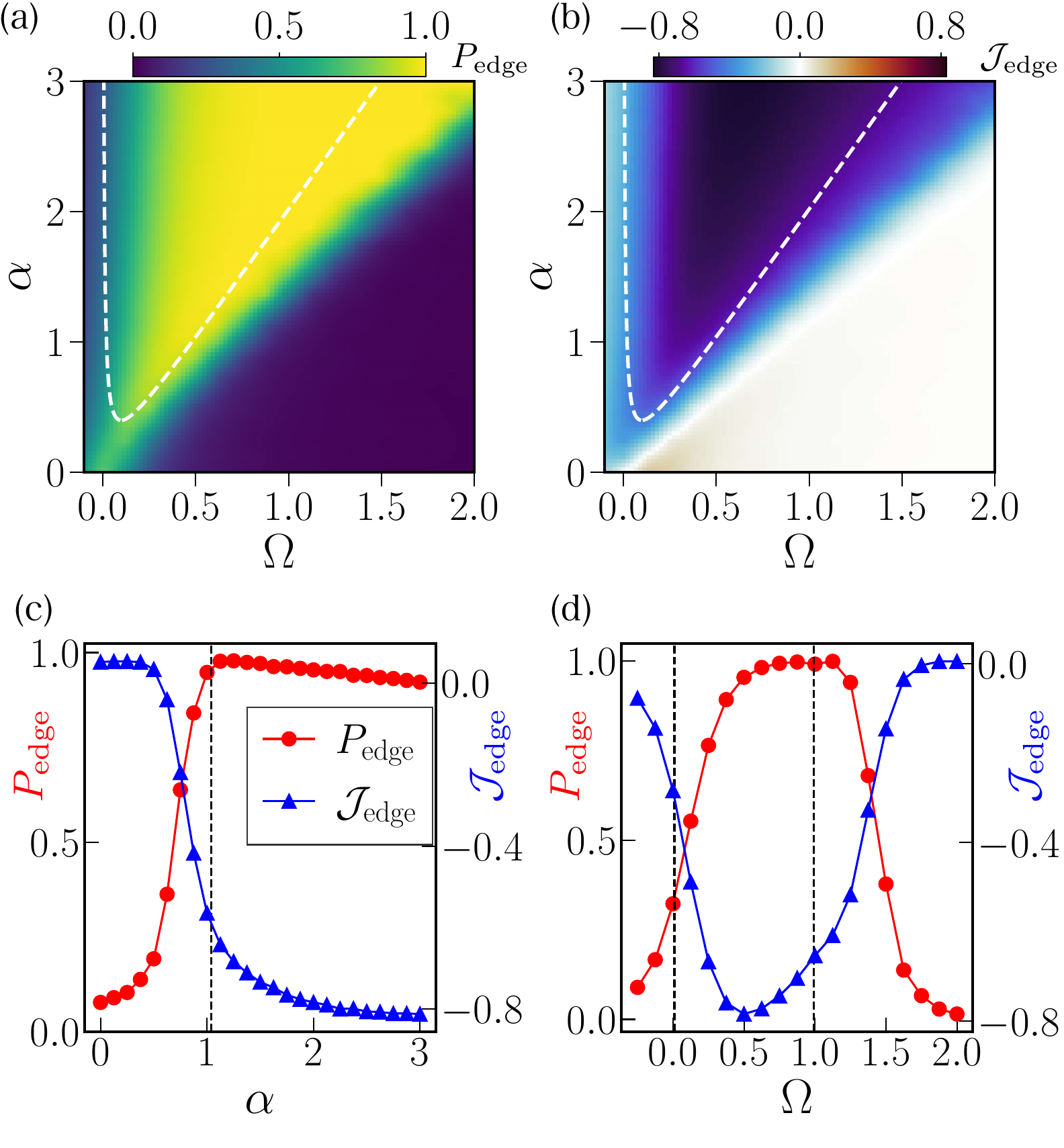}
\caption{\label{fig:PD_single} Phase diagram for a single particle confined to a circular region.
(a)~Color map of the edge occupation probability $P_\RM{edge} = \RM{Prob}[\bm r\in\RM{edge}]$ in the $\Omega$-$\alpha$ plane.
(b)~Color map of the total edge current $\mathcal J_\RM{edge}$ in the $\Omega$-$\alpha$ plane.
The white dashed lines indicate the theoretical transition line $\mathcal M=0$ [see Eq.~\eqref{eq:effective_mobility}].
(c)~Edge occupation probability and total edge current as functions of $\alpha$ at fixed $\Omega = 0.5$.
(d)~Edge occupation probability and total edge current as functions of $\Omega$ at fixed $\alpha = 2$.
The vertical dashed lines indicate the theoretical line $\mathcal M=0$.
}
\end{figure}
\section{Single-particle behavior in the presence of walls and defects} 
\label{sec:single-particle}
We begin with studying the single-particle dynamics in the presence of walls and defects.
Here, we demonstrate that the model shows edge currents that are immune to defects even in the case of a single particle. 
Throughout this section, the particle index $j$ in Eqs.~\eqref{eom1} and \eqref{eom2} is omitted. 
\subsection{Numerical simulations}
\label{ssec:simulation-single}
We first perform numerical simulations for a single particle in the presence of walls and defects.
The interaction between the particle and the walls or defects is assumed to be $\bm F = - \nabla U_\RM{wall}(\bm r)$, where $U_\RM{wall}(\bm r)$ is the wall potential.
To model the walls and defects, we use the harmonic potential given by $U_\RM{wall}(\bm r) = \epsilon[1-d(\bm r)/d_0]^2/2$
for $d(\bm r) \leq d_0$ and $U_\RM{wall}(\bm r) = 0$ for $d(\bm r) > d_0$, where $d(\bm r)$ is the distance between the particle and the walls.
We choose $d_0$ and $\tau = d_0/v_0$ as the units of length and time, respectively.
The dimensionless parameters are then
$\epsilon\mu/(d_0 v_0)$, $\Omega d_0/v_0$, $D_\RM{r}d_0/v_0$, and $\alpha d_0/\mu$.
Here, we fix $\epsilon\mu/(\sigma v_0) = 10^2$.
In the following, all quantities are expressed in these units when referring to the numerical data. 
We numerically integrate Eqs.~\eqref{eom1} and \eqref{eom2} with the Euler--Maruyama method with a discretized time step $\Delta t = 1\times 10^{-3}$.

As mentioned in the previous section, the chirality of the particle motion can change when it collides with a wall owing to the additional torque $T_j(t)$ defined in Eq.~\eqref{eq:def-torque}.
Figure~\ref{fig:model}(b) demonstrates that this chirality switching leads to an edge current.
In the simulation, the particle is confined in a square box and is initially placed at the center of the box.
The color represents the sign of the effective torque $\Omega'(t) = \Omega + T(t)$, showing that it alternates between positive and negative near the wall.

To examine the robustness of the edge current induced by chirality switching, we further show in Figs.~\ref{fig:histogram}(a)-(c) the spatial distribution function, or histogram, of a particle and the density currents for three different wall and defect geometries.
In Fig.~\ref{fig:histogram}(a), the particle is confined in a circular region and exhibits a clockwise current localized near the edge.
Figures~\ref{fig:histogram}(b) and (c) show the cases of a complex boundary and two defects with different shapes under periodic boundary conditions, respectively (see also Video~S1).
In all cases, the spatial distribution is localized near the walls and defects, and edge currents emerge.
Note that a particle that rotates clockwise (counterclockwise) in the bulk develops a current in the counterclockwise (clockwise) direction along the wall.
Furthermore, the boundaries of the defects host edge currents that flow in the same direction as the chirality of the particle.
For comparison, we also show the case of $\alpha = 0$ in Figs.~\ref{fig:histogram}(d)-(f), where neither localization of the distribution nor edge currents are observed (see also Video~S2).
These results suggest that the edge currents in this system are robust against changes in the confinement geometry and the presence of defects.

To further characterize edge localization and edge currents, we calculate the edge occupation probability $P_\RM{edge} = \RM{Prob}[\bm r\in\RM{edge}]$ and the total edge current defined by
\begin{equation}
\mathcal J_\RM{edge} = \int_\RM{edge}\dd[2]\bm r\ \expval{\bm J(\bm r)\cdot \bm e_\RM{edge}},
\end{equation}
where $\bm e_\RM{edge}$ is the unit vector parallel to the edge, $\bm J(\bm r)$ denotes the density current, and $\expval{\cdots}$ represents the ensemble average.
We use the circular confinement shown in Figs.~\ref{fig:histogram}(a) and (d), and define the edge region as $\{\bm r : |\bm r|\geq R_\RM{wall} -d_0\}$, where $R_\RM{wall}$ is the radius of the confinement.
Figures~\ref{fig:PD_single}(a) and (b) show color maps of $P_\RM{edge}$ and $\mathcal J_\RM{edge}$ in the $\Omega$-$\alpha$ plane.
The dependence of $P_\RM{edge}$ and $\mathcal J_\RM{edge}$ on $\alpha$ at $\Omega = 0.5$ and on $\Omega$ at $\alpha = 2$ is plotted in Figs.~\ref{fig:PD_single}(c) and (d), respectively (see Appendix~\ref{ape:effects_of_Dr} for the effects of $D_\RM{r}$).
For small $\alpha$, both $P_\RM{edge}$ and $\mathcal J_\RM{edge}$ are close to zero and gradually increase with increasing $\alpha$. 
As shown in Fig.~\ref{fig:model}(b), the combination $\Omega' = \Omega + T$ enables chirality switching of a particle.
For small values of $\alpha$, the interaction with the wall is not strong enough to switch the chirality of the particle.
Increasing $\Omega$ at fixed $\alpha$ shows a similar tendency.
However, both $P_\RM{edge}$ and $\mathcal J_\RM{edge}$ begin to decrease as $\Omega$ is further increased. 
These results indicate that edge localization and robust edge currents emerge in a certain parameter region, which broadens with increasing $\alpha$.
As we will discuss in Sec.~\ref{sec:topo}, the robustness of the edge currents can be understood in terms of topology.

Although our idealized minimal model combines an effective torque with isotropic interactions for simplicity, the crucial mechanism of chirality switching can be engineered easily into anisotropic chiral particles and swimmers~\cite{Kummel2013PRL, vanTeeffelen2009SoftMat}.
As a demonstration, in Appendix~\ref{ape:experiment}, we present a simple experimental realization of a ``dry'' anisotropic chiral active walker that switches chirality upon contacting a wall, and consequently exhibits a robust edge current.

\subsection{Theory}
\label{ssec:theory-single}
To theoretically describe the density localization near the walls and the edge currents, we employ a simple diffusion equation with odd terms. 
In Appendix~\ref{ape:current_single}, we derive the continuity equation $\p_t\rho(\bm r,t) = -\nabla\cdot \bm J(\bm r,t)$ with an approximated expression for the density current
\begin{equation}
\bm J \simeq - \mathcal D\nabla\rho - \mathcal D_\RM{o}\nabla^\perp\rho - \mathcal M\rho\nabla U_\RM{wall} 
- \mathcal M_\RM{o}\rho\nabla^\perp U_\RM{wall}, \label{eq:current_single}
\end{equation}
starting from the Smoluchowski equation. 
Here, $\mathcal D = v_0^2 D_\RM{r}/[2\qty(D_\RM{r}^2 + \Omega^2)]$ and $\mathcal D_\RM{o} = -{v_0^2 \Omega}/[2\qty(D_\RM{r}^2 + \Omega^2)]$ are identical to the diffusion coefficients of a free chiral active Brownian particle and $\mathcal D_\RM{o}$ represents the odd diffusivity associated with the perpendicular gradient $\nabla^\perp = (\p_y, -\p_x)$.
The last two terms describe the interaction with the wall, and the respective mobilities are given by
\begin{align}
\mathcal M:= \mu - \frac{v_0\alpha\Omega}{2\qty(D_\RM{r}^2 + \Omega^2)},
\ \ \ \ 
\mathcal M_\RM{o}:=\frac{v_0\alpha D_\RM{r}}{2\qty(D_\RM{r}^2 + \Omega^2)}. \label{eq:effective_mobility}
\end{align}
When $\alpha = 0$, $\mathcal M=\mu$ and $\mathcal M_\RM{o} = 0$.
However, for finite $\alpha$, the mobility $\mathcal M$ can become negative, and the odd mobility $\mathcal M_\RM{o}$ remains finite.
The negative mobility and finite odd mobility are responsible for density localization near the wall and edge currents, respectively.

To illustrate the combined effect of the mobilities, we consider an isotropic wall potential $U_\RM{wall}(r)$.
When the steady-state current perpendicular to the wall vanishes, the steady-state density profile can be obtained through solving
\begin{equation}
\mathcal D\dv{\rho(r)}{r} + \mathcal M \rho(r)\dv{ U_\RM{wall}(r)}{r} = 0,
\end{equation}
leading to the Boltzmann-like distribution
\begin{equation}
\rho(r) = A \exp[- \frac{\mathcal M}{\mathcal D}U_\RM{wall}(r)]. \label{eq:dist_single}
\end{equation}
Here, $A$ is the normalization factor.
When $\mathcal M$ is negative, this equation indicates that the density accumulates near the wall.
The condition for localization is therefore given by
\begin{equation}
\alpha > \alpha_\RM{c} := \frac{2\mu(D_\RM{r}^2 + \Omega^2)}{v_0\Omega}
\end{equation}
or
\begin{equation}
\Omega_\RM{c}^{(-)} < \Omega < \Omega_\RM{c}^{(+)}
\end{equation}
with $\Omega_\RM{c}^{(\pm)} := (v_0\alpha \pm \sqrt{v_0^2 \alpha^2 - 16\mu^2 D_\RM{r}^2})/(4\mu)$. 
The white dashed line in Figs.~\ref{fig:PD_single}(a) and (b) shows the theoretical prediction for the onset of edge localization, obtained from $\mathcal M = 0$.
The critical values $\alpha_\RM{c}$ and $\Omega_\RM{c}^{(\pm)}$ are also indicated by the black dashed lines in Figs.~\ref{fig:PD_single}(c) and (d), respectively.
The theory accurately captures the behavior of the numerically calculated $P_\RM{edge}$ and $\mathcal J_\RM{edge}$.

When $\rho(r)$ is localized near the edge, edge currents naturally arise owing to the odd diffusivity and odd mobility.
For simplicity, we continue to consider the isotropic case.
The tangential component of the density current is then defined as $J_\perp(r) := \bm J(r) \cdot \bm e_\theta $.
From Eq.~\eqref{eq:current_single}, we obtain
\begin{align}
J_\perp(r) = \mathcal D_\RM{o}\p_r\rho(r) + \mathcal M_\RM{o}\rho(r)\p_r U_\RM{wall}(r).
\end{align}
When the distribution $\rho(r)$ is given by Eq.~\eqref{eq:dist_single}, $J_\perp(r)$ can be written as
\begin{equation}
J_\perp(r) = \rho(r) V_\perp(r)    
\end{equation}
with $V_\perp(r) = ( \mathcal M_\RM{o} - \mathcal D_\RM{o}\mathcal M /\mathcal D)\p_r U_\RM{wall}(r)$.
Since $\mathcal D_\RM{o}$ is negative for any $\Omega>0$, $V_\perp(r)$ is positive when $\mathcal M>0$.
When the condition for edge localization is satisfied, i.e., $\mathcal M<0$, $V_\perp(r)$ can become negative, explaining the clockwise edge currents observed in the numerical simulations [see Fig.~\ref{fig:histogram}(a)].

We close this section by commenting on the quantitative discrepancies between the numerical data and the theoretical prediction.
As is well established, an active Brownian particle without chirality confined by a potential accumulates near the boundary when the persistence time $\tau_\RM{p}=1/D_\RM{r}$ is sufficiently large~\cite{Pototsky2012EPL, Elgeti2013EPL, Caprini2018SoftMat}.
Theoretically, this effect is described by a deviation from the Boltzmann distribution~\cite{Pototsky2012EPL, Fodor2016PRL, Malakar2020PRE, Chaudhuri2021JStat, Kafri2026}, written as $\rho(r) = \rho^{(0)}(r)\qty[1+\tau_\RM{p}\Delta(r) + O(\tau_\RM{p}^2)]$, where $\rho^{(0)}(r)$ is the Boltzmann distribution given by Eq.~\eqref{eq:dist_single}.
For example, when $\alpha=0$, $\Omega=0$, and $U_\RM{wall}\propto r^2$, the first-order correction is known to be $\Delta(r) = a_2 r^2 + a_4 r^4$~\cite{Malakar2020PRE, Kafri2026}.
In the present case, such a correction also exists and originates from the higher-order gradient terms neglected in deriving Eq.~\eqref{eq:current_single}.
Thus, incorporating the correction $\Delta(r)$ may improve the quantitative agreement with the numerical data, which we leave for future work.

\section{Collective behavior}
\label{sec:collective_behavior}
\begin{figure*}[t]
\centering
\includegraphics[width=18cm]{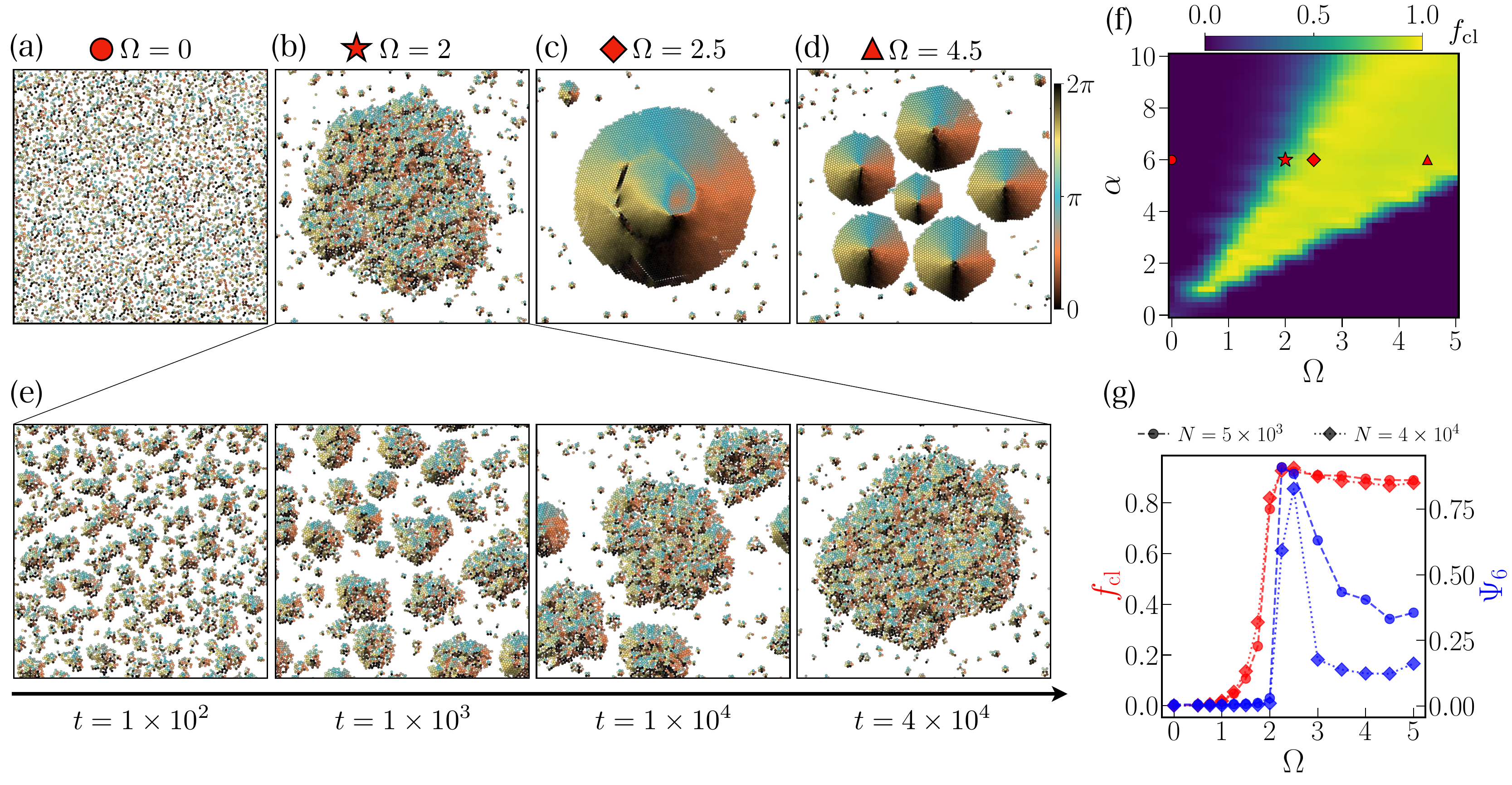}
\caption{\label{fig:config}
Phase separation in the presence of chirality switching. Typical particle configurations at (a)~$\Omega=0$, (b)~$\Omega=2$, (c)~$\Omega=2.5$, and (d)~$\Omega=4.5$ for fixed $\alpha=6$.
Each particle is colored according to the angle of its velocity with respect to the $x$-axis.
(e)~Coarsening dynamics at $\Omega=2$ and $\alpha=6$.
(f)~Phase diagram in the $\Omega$-$\alpha$ plane.
The color represents the cluster fraction.
The red symbols mark the points in the phase diagram corresponding to the snapshots shown in panels~(a)-(d).
(g)~Cluster fraction and hexatic order parameter as functions of $\Omega$ at $\alpha=6$ for two different system sizes.
The circles and diamonds represent the data for $N=5\times 10^3$ and $N=4\times 10^4$, respectively.
In all panels, the other parameters are set to $\varphi=0.4$ and $D_\RM{r}=0.1$.
Except for panel~(g), the number of particles is fixed at $N=5\times 10^3$.
The simulations are initiated from a circular droplet configuration, except for those shown in panel~(e), for which the initial configuration is a square lattice.
}
\end{figure*}
\begin{figure}[t]
\centering
\includegraphics[width=8.5cm]{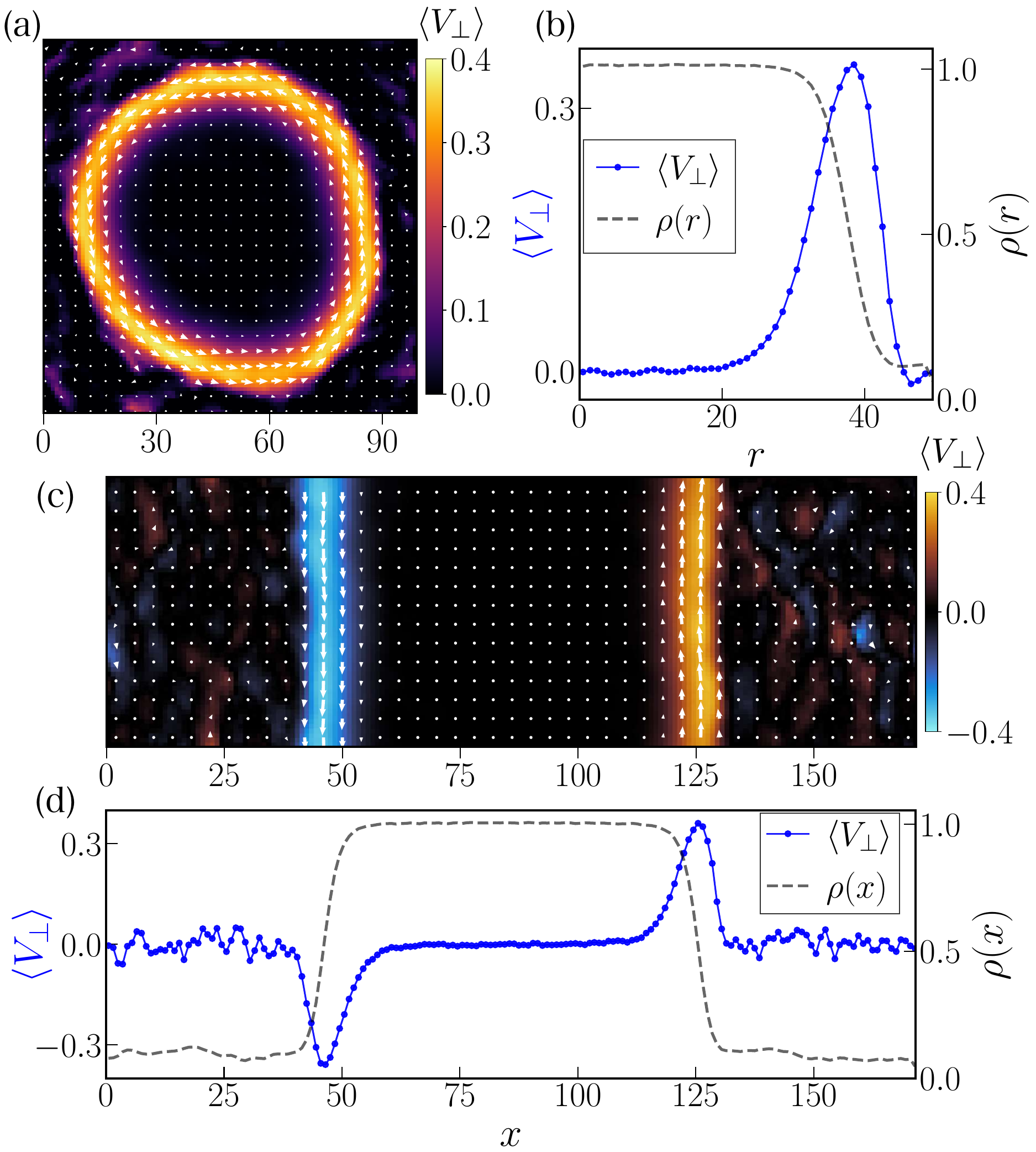}
\caption{\label{fig:collective_current}
Collective edge currents in phase-separated states.
(a,c)~Average spatial profiles of the tangential velocity component in circular and slab geometries, respectively.
(b,d)~Average tangential velocity $\expval{V_\perp}$ as a function of the distance from the center of the cluster in circular and slab geometries, respectively.
The parameters are set to $\alpha=6$, $\Omega=2$, $\varphi=0.4$, $D_\RM{r}=0.1$, and $N=5\times 10^3$.
The system aspect ratio is $L_x:L_y=1:1$ in panels~(a) and (b), and $L_x:L_y=3:1$ in panels~(c) and (d).
}
\end{figure}

In the previous section, we showed that a single particle in our model exhibits edge currents that are robust against the confinement geometry and defects.
We now turn to interacting particles and examine the effects of chirality switching on the collective behavior.
We demonstrate that the interacting particles exhibit phase separation accompanied by edge currents along interfaces.
Using an effective hydrodynamic description, we qualitatively show that phase separation with edge currents also stems from chirality switching.

\subsection{Numerical simulations}
We simulate $N$ interacting particles that obey Eqs.~\eqref{eom1} and \eqref{eom2} in a simulation box of size $L_x\times L_y$ with periodic boundary conditions.
For the pairwise interaction, we employ the harmonic potential given by
\begin{equation}
U(r) = \frac{\epsilon}{2}\qty(
1-\frac{r}{\sigma}
)^2
\end{equation}
for $r < \sigma$ and $U(r)=0$ otherwise, where $\sigma$ stands for the diameter of a particle.
We choose $\sigma$ and $\tau = \sigma/v_0$ as units of length and time. 
The dimensionless parameters are then
$\epsilon\mu/(\sigma v_0)$, $\Omega\sigma/v_0$, $D_\RM{r}\sigma/v_0$, $\alpha\sigma/\mu$, and the packing fraction $\varphi = \pi\sigma^2N/(4L_x L_y)$.
Here, we fix $\epsilon\mu/(\sigma v_0) = 10^2$ and $D_\RM{r}\sigma/v_0 = 0.1$.
Throughout this subsection, we henceforth denote all quantities in the aforementioned units unless otherwise noted. 
We numerically integrate Eqs.~\eqref{eom1} and \eqref{eom2} with the Euler--Maruyama method with a discretized time step $\Delta t = 1\times 10^{-3}$. 

In Figs.~\ref{fig:config}(a)-(d), we demonstrate that the system undergoes clustering when both $\alpha$ and $\Omega$ are finite.
In each panel, the color represents the angle of the velocity $\dot{\bm r}_j$ with respect to the $x$-axis.
At $\Omega=2$, the cluster exhibits a liquid-like structure, whereas at larger values of $\Omega$ clusters are strongly ordered and undergo body rotation (see also Videos~S3 and S4).
Figure~\ref{fig:config}(e) shows the time evolution of the system at $\Omega=2$ and $\alpha=6$ initiated from the square-lattice initial configuration (see also Video~S5).  
One observes spinodal-like coarsening behavior that resembles standard liquid-vapor phase separation.

The phase behavior in the parameter space of $\Omega$-$\alpha$ is summarized in Fig.~\ref{fig:config}(f).
The color represents the cluster fraction $f_\RM{cl} = \expval{N_\RM{cl}}/N$. 
$N_\RM{cl}$ denotes the number of particles belonging to a cluster, calculated by $N_\RM{cl} = \sum_{j=1}^{N}\Theta(\varphi_j - \varphi_\RM{th})$, where 
$\varphi_j$ is local packing fraction around the $j$-th particle, $\varphi_\RM{th}$ is threshold value, which we set to $\varphi_\RM{th} = 0.7$, and $\Theta(\cdot)$ is the Heaviside step function.
Note that the clustering observed here is distinct from motility-induced phase separation, which takes place in self-propelled particles with large persistence length~\cite{Fily2012PRL, Bialk2013EPL, Buttinoni2013PRL, Cates2015AnnuRev}.
Indeed, the present system remains homogeneous at $\Omega =\alpha = 0$, and the cluster formation and phase separation can therefore be attributed to chirality switching (see also Appendix~\ref{ape_ssec:polarization} for another distinction from MIPS). 
This point is clarified by an effective hydrodynamic description discussed in the next section.
In Fig.~\ref{fig:config}(g), we show the cluster fraction as a function of $\Omega$, as well as the hexatic order parameter defined by $\Psi_6 = |\sum_{j=1}^N\psi_6(\bm r_j)/N|$ with $\psi_6(\bm r_j) = \sum_{k\in\mathcal N_6(\bm r_j)}e^{6i\theta_{jk}}/6$. 
Here, $\mathcal N_6(\bm r_j)$ denotes the set of six nearest neighbors of the $j$-th particle, and $\theta_{jk}$ is the angle between line $|\bm r_j-\bm r_k|$ and the $x$-axis.
As $\Omega$ increases, both the cluster fraction and the hexatic order parameter increase.
However, the values of $\Omega$ at which $f_\RM{cl}$ and $\Psi_6$ jump do not coincide.
One first observes an increase in $f_\RM{cl}$, followed by the growth of the hexatic order.
We also confirm that this trend is quantitatively the same for two different system sizes, $N=5\times 10^3$ and $N=4\times 10^4$.
The point at which $f_\RM{cl}\simeq 0.8$ and $\Psi_6\simeq 0$ corresponds to the phase-separated state with edge currents shown in Fig.~\ref{fig:collective_current}.
In contrast, the state in which the entire cluster rotates, as shown in Figs.~\ref{fig:config}(c) and (d), is characterized by a finite cluster fraction and a finite hexatic order parameter.
Note that the decrease in $\Psi_6$ for $\Omega > 2.5$ stems from the disintegration of the clusters.
The hexatic order remains finite within each cluster (see Appendix~\ref{ape_ssec:hexatic} for $\Psi_6$ calculated for the largest cluster).
These rotating clusters with hexatic structures are reminiscent of chiral crystals exhibiting odd dynamics~\cite{Yan2015SoftMat, Petroff2015PRL, Tan2022Nature, Bililign2022NatPhys, Huang2025PNAS}. 
Since odd terms naturally arise in the coarse-grained description of our model, as discussed below, we speculate that the rotation in this regime originates from the emergence of odd elasticity~\cite{Fruchart2023AnnuRev}, although this issue is beyond the scope of the present study.

In Fig.~\ref{fig:collective_current}, we show that the phase-separated state displays steady currents along the interface. 
Figure~\ref{fig:collective_current}(a) shows the spatial distribution of the time-averaged tangential velocity $\expval{V_\perp(\bm r)}$, which is calculated as the inner product of $\dot{\bm r}_j$ and the unit vector perpendicular to $\bm r_j - \bm r_\RM{G}$, where $\bm r_\RM{G}$ is the center-of-mass position.
The tangential velocity as a function of the distance from the center of the cluster, together with the projected density profile, is shown in Fig.~\ref{fig:collective_current}(b).
One can clearly see that $\expval{V_\perp}$ has a sharp peak at the interface position.
In Figs.~\ref{fig:collective_current}(c) and (d), we further show the emergence of edge currents in a phase-separated state with slab geometry, where the cluster has a band shape.
Note that the behavior of the edge currents in this phase-separated state is insensitive to changes in the global density, as shown in Appendix~\ref{ape_ssec:density_dependence}.
As will be shown in Sec.~\ref{sec:topo}, this phase separation turns out to correspond to the coexistence of two topologically distinct domains, and the edge currents are therefore topological edge modes.

\subsection{Hydrodynamic theory} 
\subsubsection{Effective hydrodynamic equations}

\begin{figure*}[t]
\centering
\includegraphics[width=16cm]{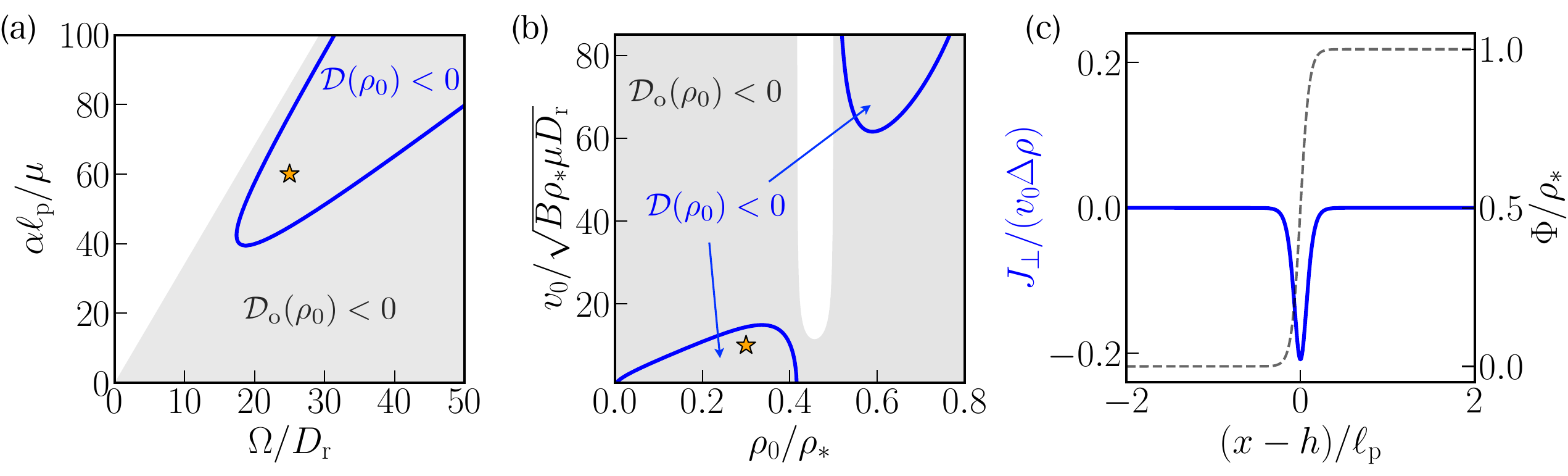}
\caption{\label{fig:instability}
Effective hydrodynamic theory. (a)~Linearly unstable region obtained from $\mathcal D(\rho_0)<0$ (inside the blue line).
The gray shaded region indicates where the odd diffusivity $\mathcal D_\RM{o}(\rho_0)$ is negative.
The horizontal and vertical axes are nondimensionalized by $D_\RM{r}$ and $\mu/\ell_\RM{p}$, respectively, where $\ell_\RM{p} = v_0/D_\RM{r}$.
The other dimensionless parameters are set to $v_0/\sqrt{B\rho_*\mu D_\RM{r}} = 10$ and $\rho_0/\rho_* = 0.3$, where $\rho_* = v_0/(\mu\gamma)$ denotes the density at which the effective self-propulsion speed $v(\rho_0)$ vanishes.
(b)~Linearly unstable region at fixed $\alpha\ell_\RM{p}/\mu = 60$ and $\Omega/D_\RM{r} = 25$.
The stars in panels~(a) and (b) indicate the same point in parameter space.
(c)~Illustration of the edge current in a phase-separated state with a flat interface.
The blue line depicts Eq.~\eqref{eq:edge_current}, and the black dashed line represents the density profile given by the sigmoid function (see the main text).
The parameters are set to $\Omega/D_\RM{r}=25$, $\alpha\ell_\RM{p}/\mu =60$, $v_0/\sqrt{B\rho_*\mu D_\RM{r}} = 10$, $\rho_0/\rho_* = 0.3$, $\rho_\RM{l}/\rho_*=1$, $\rho_\RM{g}/\rho_*=0$, and $\xi/\ell_\RM{p} = 0.1$.
}
\end{figure*}

To obtain further insights, we develop an effective hydrodynamic description for our particle model that qualitatively accounts for the phase separation accompanied by edge currents observed in the numerical simulations above.
Several approaches have been proposed for deriving hydrodynamic equations for active particles, some of which have been applied to chiral active particles~\cite{Ma2022JCP, Bickmann2022JCP, Sansa2022JCP, Kuroda2023JStatMech, Kalz2024JPhysA, Marconi2026NewJPhys, Maire2026}.
One established approach is to make a closure approximation for the BBGKY equations corresponding to the particle model under consideration~\cite{Bialk2013EPL, Speck2015JCP}.
Here, we follow this standard procedure.
The details of the derivation are described in Appendix~\ref{ape:hydro}.

Our starting point is the equation for the one-body distribution function, denoted by $\Psi = \Psi(\bm r,\phi,t)$, which has a hierarchical structure.
In Appendix~\ref{ape_ssec:closure}, we close the hierarchy by making several approximations for the interaction terms and then obtain
\begin{equation}
\begin{split}
\p_t \Psi = &-\nabla\cdot \Psi\qty[v(\rho)\bm e(\phi) - \mu B\nabla\rho] \\
&-\p_\phi\qty[\Omega(\rho) - B\alpha \qty(\bm e(\phi) \cdot \nabla \rho) -D_\RM{r} \p_\phi]\Psi, \label{eq:one-body-2}
\end{split}
\end{equation}
where we define density-dependent self-propelled speed and chiral torque as
\begin{equation}
v(\rho) := v_0 - \mu \gamma \rho, \ \ \ \ \   
\Omega(\rho) :=  \Omega - \alpha \gamma \rho,
\end{equation}
respectively. 
The coefficients $\gamma$ and $B$ are determined by the pairwise potential and the pair distribution function, and are positive.
Our purpose here is to obtain equations for the density field $\rho(\bm r,t)=\expval*{\sum_{j=1}^{N}\delta(\bm r-\bm r_j(t))}$ and the polarization field $\bm p(\bm r,t) = \expval*{\sum_{j=1}^{N}\bm e(\phi_j(t))\delta(\bm r-\bm r_j(t))}$ from Eq.~\eqref{eq:one-body-2}.
By integrating Eq.~\eqref{eq:one-body-2} over $\phi$, we get a continuity equation $\p_t\rho(\bm r,t) = -\nabla \cdot \bm J(\bm r,t)$ with the density current being 
\begin{equation}
 \bm J(\bm r,t) = v(\rho) \bm p(\bm r,t) - D_\RM{e}(\rho) \nabla \rho(\bm r,t).
 \label{eq:density_current_1}
\end{equation}
Here, the diffusion coefficient is defined by $D_\RM{e}(\rho) = \mu B \rho(\bm r)$.
Likewise, the equation for the polarization reads
\begin{equation}
\begin{split}
\p_t \bm p = -\nabla\cdot \qty[ v(\rho)\mathsf Q +\frac{\rho}{2} v(\rho)\mathbbm 1 - \mu B\bm p \nabla\rho] - D_\RM{r}\bm p \\
- \Omega(\rho) \underline{\bm\epsilon} \bm p +  B\alpha \underline{\bm\epsilon} \qty[\mathsf Q +\frac{\rho}{2}\mathbbm 1]\cdot \nabla\rho, \label{eq:polar-1}
\end{split}
\end{equation}
where $\mathbbm 1$ is the identity matrix, $\underline{\bm\epsilon}$ represents the Levi--Civita tensor in two dimensions whose elements are given by $\epsilon_{x,y} = - \epsilon_{y,x}=1$ and  $\epsilon_{x,x} = \epsilon_{y,y}=0$, and 
$\mathsf Q(\bm r,t)$ is the nematic tensor.
In Appendix~\ref{ape_ssec:nematic}, we find the approximated form of $\mathsf Q(\bm r,t)$ as 
\begin{equation}
\mathsf Q(\bm r) \simeq -\mathsf A \mathsf q(\bm r) \label{eq:nemtic-1}
\end{equation}
with matrices
\begin{align}
q_{\alpha,\beta} = \p_\alpha v(\rho)p_{\beta} + \p_\beta v(\rho)p_\alpha - \p_\mu v(\rho) p_\mu\delta_{\alpha,\beta} \label{eq:tensor_q}
\end{align}
and 
\begin{align}
\mathsf A = \frac{1}{32D_\RM{r}^2 + 8 \Omega^2(\rho)}
\mqty(
2D_\RM{r} & -\Omega(\rho) \\
 \Omega(\rho) & 2D_\RM{r} \label{eq:matrix_A}
).
\end{align}
Using this approximated form of $\mathsf Q(\bm r,t)$ and neglecting the terms $O(\nabla \bm p\nabla\rho)$, we end up with
\begin{equation}
\begin{split}
\p_t \bm p = &- [D_\RM{r} - \eta (\rho)\nabla^2]\bm p  - \frac{1}{2}\nabla\qty[\rho v(\rho)] \\
&-\qty[ \Omega(\rho) + \eta_\RM{o}(\rho)\nabla^2 ]\bm p^\perp 
+ \frac{1}{2}\nabla^{\perp}\qty[\rho v_\perp(\rho)] \label{eq:polar-2}
\end{split}
\end{equation}
governing the evolution of the polarization field.
Here, $\eta(\rho)$ and $\eta_\RM{o}(\rho)$ represent ``viscous'' coefficients for the polarization defined by
\begin{align}
&\eta(\rho) := \frac{v^2(\rho) D_\RM{r}}{16D_\RM{r}^2 + 4 \Omega^2(\rho)}, \\ 
&\eta_\RM{o} (\rho) :=\frac{v^2(\rho) \Omega(\rho)}{32D_\RM{r}^2 + 8 \Omega^2(\rho)}. \label{eq:odd_viscous1}
\end{align}
In Eq.~\eqref{eq:polar-2}, we also define $v_\perp(\rho) := \alpha B \rho/2$. 
The last three terms involving $\bm p^\perp=(p_y,-p_x)$ and $\nabla^\perp =(\p_y,-\p_x)$ in Eq.~\eqref{eq:polar-2} originate from the chiral terms in Eq.~\eqref{eom2}.
In particular, the additional torque responsible for chirality switching, Eq.~\eqref{eq:def-torque}, gives rise to the density dependence of the effective torque $\Omega(\rho)$ and the perpendicular density gradient $\nabla^\perp\rho$.

\subsubsection{Instability and edge currents}

We now discuss the instability and edge currents that originate from chirality switching using the equations obtained above.
Although one can perform a linear stability analysis of the coupled equations for the density and polarization, we instead work with the simplified equation for the density field alone.
See Appendix~\ref{ape_ssec:stability} for the linear stability analysis of the coupled equations.
Since $\bm p$ is non-conserved and its evolution has a damping term, the spatiotemporal behavior of $\bm p$ relaxes faster than the density field $\rho$.
We thus adopt the adiabatic approximation through $\p_t\bm p\simeq 0$, leading to the quasi-stationary solution
\begin{equation}
\bm p\simeq -\frac{1}{2}\mathsf W^{-1}(\rho)\nabla[\rho v(\rho)] +\frac{1}{2}\mathsf W^{-1}(\rho) \nabla^\perp[\rho v_\perp(\rho)] +O(\nabla^3\rho),
\end{equation}
where
\begin{equation}
\mathsf W(\rho) = \mqty(D_\RM{r} & \Omega(\rho) \\ -\Omega(\rho) & D_\RM{r}).    
\end{equation}
Plugging this adiabatic solution into Eq.~\eqref{eq:density_current_1}, one obtains the density current with odd diffusivity:
\begin{equation}
\bm J(\bm r,t) \simeq - \mathcal D(\rho) \nabla\rho(\bm r,t) - \mathcal D_\RM{o}(\rho) \nabla^\perp\rho(\bm r,t), \label{eq:current-many1}
\end{equation}
where 
\begin{align}
&\mathcal D(\rho):= D_\RM{e}(\rho) + \frac{v(\rho)\qty[D_\RM{r}(v_0-2\mu\gamma\rho) - \alpha B\rho\Omega(\rho)]}{2[D_\RM{r}^2 + \Omega^2(\rho)]}, \label{eq:collective_diffusion} \\ 
&\mathcal D_\RM{o}(\rho) := -\frac{v(\rho)[D_\RM{r}\alpha B \rho + (v_0-2\mu\gamma\rho)\Omega(\rho)]}{2[D_\RM{r}^2 + \Omega^2(\rho)]}.
\end{align}
Since the odd diffusion term is divergence-free, it does not affect the density field and the evolution of the density profile is determined by the diffusion equation 
\begin{equation}
\p_t\rho(\bm r,t) = \nabla\cdot \mathcal D(\rho)\nabla\rho(\bm r,t). \label{eq:diffusion_adiabatic}
\end{equation}
The homogeneous solution $\rho = \rho_0$ is linearly unstable if $\mathcal D(\rho_0)<0$. 

Importantly, there are two mechanisms that can render the diffusion coefficient $\mathcal D(\rho_0)$ negative.
One is the density dependence of the effective self-propulsion speed $v(\rho_0)$, which gives rise to the well-known motility-induced phase separation (MIPS)~\cite{Bialk2013EPL, Speck2015JCP}.
The other is the density dependence of the effective torque $\Omega(\rho_0)$.
As is clear from the definition of $\Omega(\rho)$, $\mathcal D(\rho_0)$ can become negative as $\alpha$ increases.
In Fig.~\ref{fig:instability}(a), we show the unstable region in the $\Omega$-$\alpha$ plane obtained from the condition $\mathcal D(\rho_0)<0$.
Figure~\ref{fig:instability}(b) further shows that, in the $\rho$-$v_0$ plane at fixed values of $\alpha$ and $\Omega$, two separate unstable regions exist.
These two unstable regions remain separated as long as $\mathcal D(\rho_0=\rho_{*}/2)>0$.
The unstable region at large $v_0$ corresponds to MIPS~\cite{Bialk2013EPL, Speck2015JCP}, whereas the lower unstable region corresponds to that shown in Fig.~\ref{fig:instability}(a).
This latter linear instability arises from the density dependence of $\Omega(\rho_0)$ and we argue that it underlies the phase separation observed in the numerical simulations.
We note that, as shown in Appendix~\ref{ape_ssec:stability}, there is another type of instability that sets in at a finite wavenumber and thus corresponds to states with multiple finite-sized clusters as reported in Ref.~\cite{Ma2022JCP}. 

\begin{figure*}[t]
\centering
\includegraphics[width=1.0\linewidth]{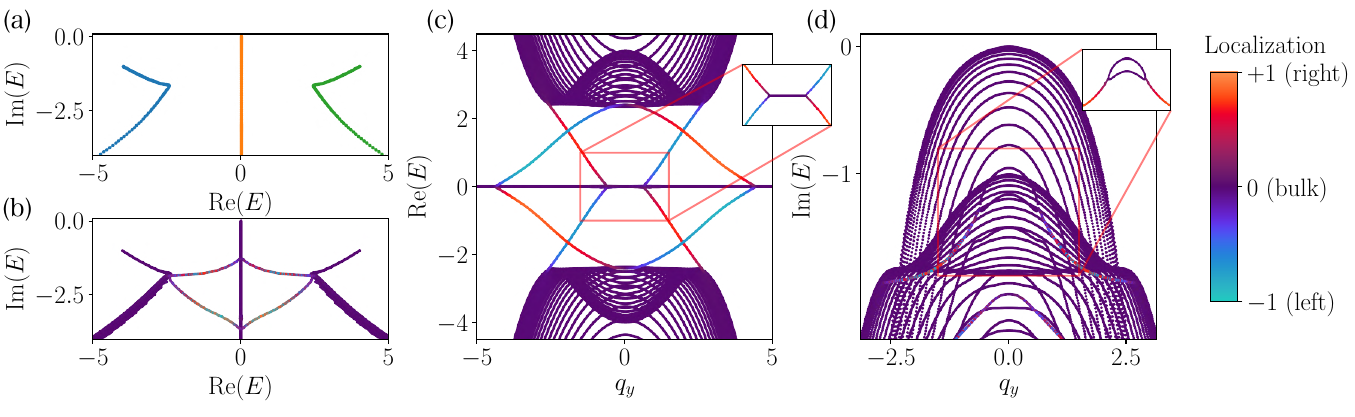}
\caption{\label{fig:band}Band structure obtained from the effective Hamiltonian.
(a)~Energy bands under half-periodic boundary conditions.
(b)-(d)~Energy bands for systems with periodic boundary conditions in the $y$-direction and open boundary conditions in the $x$-direction.
(b)~Real and imaginary parts of the energy eigenvalues.
(c,d)~Real and imaginary parts of the energy eigenvalues plotted as functions of $q_y$, respectively.
The color indicates the degree of edge localization of the eigenvectors.
A value of $+1$ ($-1$) indicates that an eigenvector is completely localized at the right (left) edge.
The insets show magnifications of the edge modes enclosed by red rectangles. 
The parameters are set to $b=0.1$, $v_\RM{s} = 1$, $\Lambda_\perp = 0.1$, $\nu = 0.25$, $\nu_\RM{o}=0.5$, and $\mathcal W=4$. 
}
\end{figure*}

Although the odd diffusion term does not affect the density profile, one can readily see that it is responsible for the emergence of edge currents when the system is phase-separated.
As a simple example, let us consider a completely phase-separated state into gas and liquid phases with a flat interface parallel to the $y$-direction, expressed as $\rho(\bm r) = \Phi(x-h)$, where $\Phi(u)$ is a sigmoid function and $h$ denotes the constant interface height.
For example, we take $\Phi (u) = (\rho_\RM{l} +\rho_\RM{g})/2+ (\rho_\RM{l} - \rho_\RM{g})\tanh\qty({u}/{\xi})/2$ with the interface width $\xi \ll 1$.
Here, $\rho_\RM{l}$ and $\rho_\RM{g}$ denote the densities of the liquid and gas phases, respectively.
This situation corresponds to the left half of Figs.~\ref{fig:collective_current}(c) and (d).
The tangential component of the density current is then written as
\begin{equation}
J_\perp(x) = J_y(x) =  \mathcal D_\RM{o}(\rho)\p_x\Phi(x-h),
\label{eq:edge_current}
\end{equation}  
which implies that $J_\perp(x)$ is finite only near the interface position where the density varies and vanishes elsewhere.
The direction of the edge current is determined by the sign of $\mathcal D_\RM{o}$.
The linearized theory under the adiabatic approximation predicts that $\mathcal D_\RM{o}$ is negative within the unstable region induced by chirality switching, as shown in Figs.~\ref{fig:instability}(a) and (b).
In the present example, where the liquid phase is located at $x>h$, the edge current $J_\perp(x)$ flows in the negative $y$-direction for $\mathcal D_\RM{o} < 0$, as illustrated in Fig.~\ref{fig:instability}(c), which is qualitatively consistent with the numerical observations shown in Fig.~\ref{fig:collective_current}.
Note that $\mathcal D_\RM{o}(\rho_0)$ is negative even within the unstable region corresponding to MIPS, as shown in Fig.~\ref{fig:instability}(b).
This behavior also results from chirality switching.
Indeed, when $\alpha=0$, one can readily show that $\mathcal D_\RM{o}(\rho_0)$ is positive in the MIPS region, and thus the edge current flows in the positive $y$-direction, consistent with previous numerical observations~\cite{Ma2022JCP, Metzger2026}.
Although we have illustrated the emergence of edge currents only for a flat interface, this discussion can be straightforwardly extended to interfaces of arbitrary shape.

We remark that Eq.~\eqref{eq:diffusion_adiabatic} does not contain a stiffness term, and thus it cannot yield a density profile with a finite interfacial width of the type assumed above.
This issue can be resolved by taking into account higher-order gradient terms.
In Appendix~\ref{ape_ssec:field_theory}, we discuss a possible scalar field theory for our model that includes stiffness terms.


\section{Topological edge modes}
\label{sec:topo} 

In the previous two sections, we observed edge currents in both single- and many-body situations.
So far, we have explained these edge currents in terms of odd diffusivity.
However, such descriptions do not explain why the currents are robust.
Here, we show that the edge currents in our system indeed correspond to topological edge modes.

\subsection{Effective Hamiltonian and topological invariant}

To exploit a mathematical analogy between quantum systems and our chiral active system, we first linearize the hydrodynamic equations~\eqref{eq:density_current_1} and \eqref{eq:polar-2} derived in the previous section.
For notational simplicity, we also nondimensionalize the equations using the characteristic length $\ell_\RM{p} = v_0/D_\RM{r}$, the time scale $1/D_\RM{r}$, and the reference density $\rho_{*} ={v_0}/(\mu\gamma)$.
We rescale position, time, density, and polarization as $\bm r\to \bm r/\ell_\RM{p}$, $t\to t D_\RM{r}$, $\rho\to \rho/\rho_*$, and $\bm p\to \bm p/\rho_*$, respectively.
The set of small fluctuations around the stationary state in Fourier space, $\hat{\bm \psi}(\bm q) = (\delta\hat\rho,\delta \hat p_x, \delta \hat p_y)$, satisfies $i\p_t \hat{\bm \psi}(\bm q) = \mathsf H(\bm q)\hat{\bm \psi}(\bm q)$.
As shown in Appendix~\ref{ape_ssec:Hamiltonian}, the Hamiltonian-like matrix $\mathsf H(\bm q)$ can be transformed as
\begin{align}
\tilde{\mathsf H}(\bm q) = \mqty( -i b q^2 & v_\RM{s} q_x & v_\RM{s} q_y \\ 
v_\RM{s} q_x - \Lambda_\perp q_y & -i(1+\nu q^2) & -i(\mathcal W - \nu_{\rm o}q^2) \\ 
v_\RM{s} q_y + \Lambda_\perp q_x & i(\mathcal W - \nu_{\rm o}q^2) & -i(1+\nu q^2) ). \label{eq:Hamiltonian}
\end{align}
Here, all coefficients are dimensionless.
The important parameters are the effective chiral torque,
$\mathcal W = \Omega/D_\RM{r} - \alpha\ell_\RM{p}\rho_0/(\mu\rho_*)$,
and the odd viscosity,
$\nu_\RM{o} = \eta_\RM{o}(\rho_0)/(\ell_\RM{p}^2D_\RM{r})$.
See Appendix~\ref{ape_ssec:stability} for the definitions of the other dimensionless parameters.
The matrix in Eq.~\eqref{eq:Hamiltonian} is non-Hermitian, i.e., $\tilde{\mathsf H}(\bm q) \neq \tilde{\mathsf H}^\dagger(\bm q)$, but its Hermitian part is the same as that considered in Ref.~\cite{Souslov2019PRL} (see also Appendix~\ref{ape_ssec:Hamiltonian}).
We also note that the noninteracting-particle case can be described by the same equations by setting $b = \Lambda_\perp = 0$ and taking the limit $\rho_0\to 0$.

In order to identify a topological invariant, Fourier space must be compact, i.e., $\tilde{\mathsf H}(\bm q)$ must become independent of the direction of the wavevector $\bm q$ in the limit $q\to \infty$.
As shown in Refs.~\cite{Souslov2019PRL, Fujii2025SciPost}, in chiral active fluids, the odd-viscous term $\nu_\RM{o}q^2$ plays a role in compactifying Fourier space, which allows one to calculate well-defined Chern numbers.
In the Hermitian case, it was shown that the Chern numbers are determined by the signs of the chirality and the odd viscosity~\cite{Souslov2019PRL, Fujii2025SciPost}.
Even in non-Hermitian systems, one can calculate the Chern numbers as in the Hermitian case when the band is separable~\cite{Shen2018PRL}.
In Fig.~\ref{fig:band}(a), we show a typical behavior of the eigenvalues of Eq.~\eqref{eq:Hamiltonian} in the complex plane.
Clearly, each band satisfies $E_n(\bm q) \neq E_m(\bm q)$ for all $m(\neq n)$ and all $\bm q$.
We also note that the gap in this band structure is classified as a line gap, whose topological properties are known to be essentially equivalent to those of Hermitian systems~\cite{Kawabata2019PRX, Ashida2020}.
Indeed, one obtains the Chern numbers as follows (see Appendix~\ref{ape_ssec:Chern} for the derivation):
\begin{align}
\mathcal C_{\pm} = \mp\qty[ \RM{sgn}(\nu_\RM{o}) + \RM{sgn}(\mathcal W)],
\end{align} 
which is equivalent to that obtained in Refs.~\cite{Souslov2019PRL, Fujii2025SciPost}.
In our system, the odd viscosity is related to the effective chirality, as defined in Eq.~\eqref{eq:odd_viscous1}.
Therefore, the Chern number is determined solely by the effective chiral torque:
\begin{align}
\mathcal C_{\pm} = \mp 2 \RM{sgn}\qty[\Omega(\rho_0)],
\end{align}
where we return to the original dimensions.
We remark that even for noninteracting particles, the Chern numbers remain finite: $\mathcal C_{\pm} = \mp 2 \RM{sgn}\qty(\Omega)$, where $\Omega = \Omega(\rho_0\to 0)$ is the ``bare'' chiral torque in Eq.~\eqref{eom2}.
We also note that, in the limit $\Omega(\rho_0)\to 0$, the Chern numbers become ill-defined because Fourier space is no longer compact~\cite{Souslov2019PRL, Fujii2025SciPost}.

\subsection{Band structure}

Topological invariants in the bulk are usually related to the number of edge modes in systems with boundaries.
When two domains, $\RM{A}$ and $\RM{B}$, characterized by different Chern numbers, $\mathcal C_\RM{A}$ and $\mathcal C_\RM{B}$, are joined, the bulk-boundary correspondence states that the number of edge modes is given by $\abs{\mathcal C_\RM{A} - \mathcal C_\RM{B}}$.
Although there is no guarantee that the bulk-boundary correspondence holds for non-Hermitian systems, we confirm its validity in our system by examining the band structure. 

Following Ref.~\cite{Sone2020NatComm}, we replace $iq_x$ with $\p_x$ in the effective Hamiltonian, Eq.~\eqref{eq:Hamiltonian}, and impose periodic boundary conditions in the $y$-direction and open boundary conditions in the $x$-direction.
We discretize space in the $x$-direction using the central-difference method and numerically compute the eigenvalues (see Appendix~\ref{ape_ssec:BandStructure} for the technical details).
Figures~\ref{fig:band}(a) and (b) show the energy bands in the complex plane for the systems with periodic and open boundary conditions, respectively.
We also show the real parts of the eigenvalues as functions of $q_y$ in Fig.~\ref{fig:band}(c).
The color indicates the degree of localization of the eigenvectors at the edges.
One can clearly identify gapless edge modes.
Figure~\ref{fig:band}(d) shows the imaginary parts of the eigenvalues as functions of $q_y$.
All eigenvalues, including those of the edge modes, have non-positive imaginary parts, suggesting that the edge modes are linearly stable.
In Appendix~\ref{ape_ssec:EdgeModes}, we also show that, by adopting an ansatz for the eigenmodes, one can obtain edge solutions that quantitatively agree with those obtained from direct numerical calculations.

Although the bulk topological properties are equivalent to those in the Hermitian case, the band structure under open boundary conditions exhibits a non-Hermitian feature: exceptional points appear, as suggested by the insets of Figs.~\ref{fig:band}(c) and (d), which have been linked to edge modes~\cite{Sone2020NatComm}.

\subsection{Edge modes}

\begin{figure}[t]
\centering
\includegraphics[width=1.0\linewidth]{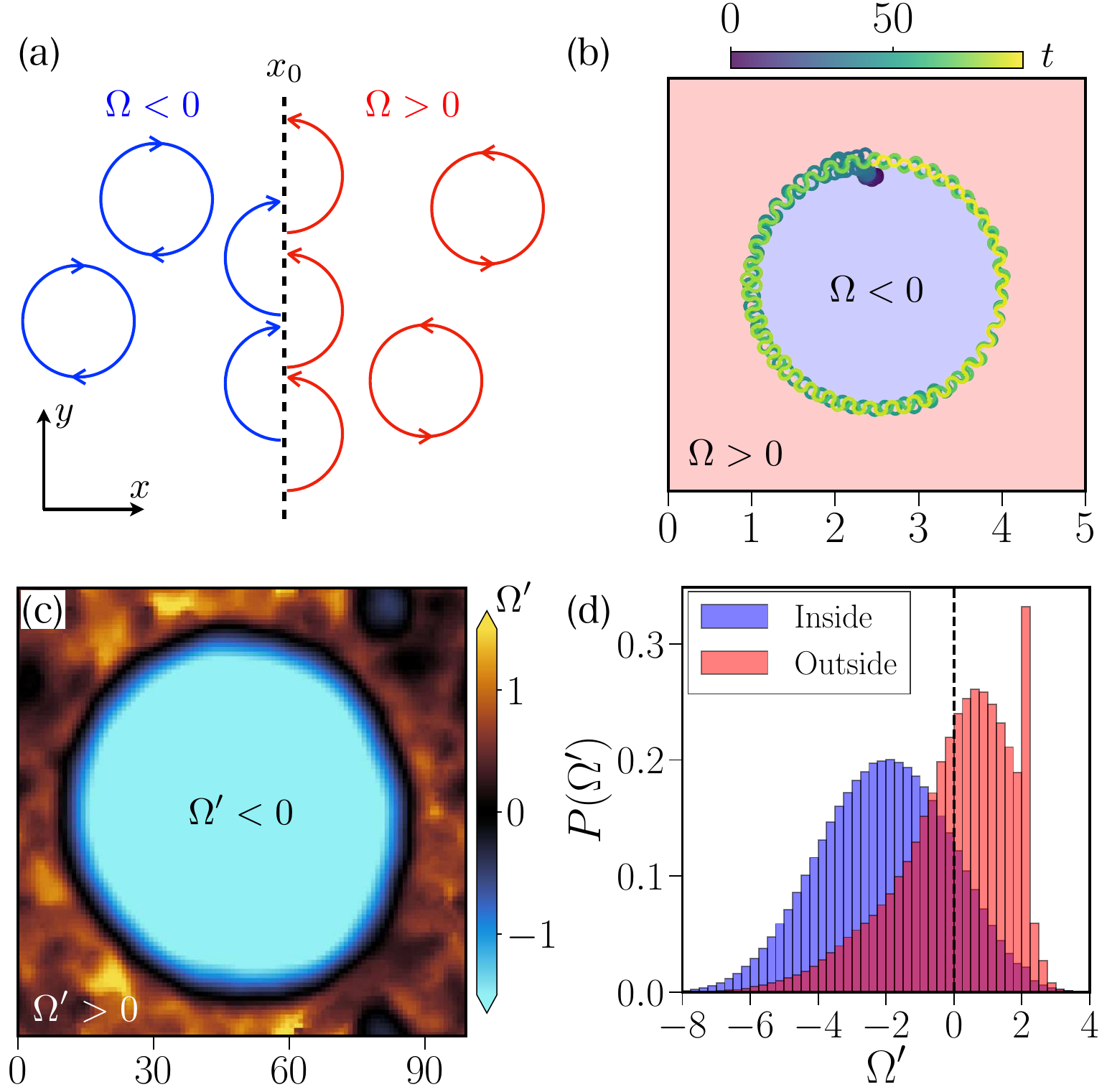}
\caption{\label{fig:torque_sgn}(a)~Schematic of an edge current along a domain wall separating regions of opposite chirality, and (b)~its numerical demonstration in a single-particle simulation.
The blue and red regions represent domains with negative and positive $\Omega$, respectively.
In the simulation, we solve Eqs.~\eqref{eom1} and \eqref{eom2} without $T_j(t)$ prescribing a torque $\Omega$ whose sign depends on the position.
The parameters are set to $|\Omega| =15$ and $D_\RM{r}=0.05$.
(c)~Color map of the spatially averaged effective chirality $\Omega' = \Omega + T_j$ in a phase-separated state.
(d)~Histogram of the effective chirality $\Omega'$ inside and outside the droplet.
The parameters are set to $\alpha=6$, $\Omega=2$, $\varphi=0.4$, $D_\RM{r}=0.1$, and $N=5\times 10^3$ (the same as those in Fig.~\ref{fig:collective_current}).
}
\end{figure}

As shown in the previous section, the edge currents in our system can be understood in terms of the bulk-boundary correspondence.
As an illustration, we first consider the single-particle case with $\alpha=0$, in which the model is identical to a chiral active Brownian particle.
We introduce two domains with $\Omega>0$ and $\Omega <0$ separated at $x=x_0$, as schematically shown in Fig.~\ref{fig:torque_sgn}(a).
As discussed in the previous section, the Chern numbers are given by $\mathcal C_\pm = \mp 2\RM{sgn}(\Omega)$, indicating that the two domains are characterized by different Chern numbers.
Thus, according to the bulk-boundary correspondence, edge modes are expected to appear at the domain wall.
In Fig.~\ref{fig:torque_sgn}(b), we numerically demonstrate the existence of a current along the domain wall.
In the simulation, we consider a circular domain with $\Omega<0$, while $\Omega>0$ outside the domain.
This behavior is also intuitively reasonable, since chirality switching across the domain wall causes the particle to move along the boundary between the two domains.
In Appendix~\ref{ape_ssec:domain}, we also demonstrate that the same phenomenon can be observed for interacting particles. 

One can now interpret the edge currents of a single particle in our model, discussed in Sec.~\ref{sec:single-particle}, from the perspective of topology.
As shown in Fig.~\ref{fig:model}(b), a particle confined by a wall potential undergoes chirality switching when it hits the wall due to the additional torque defined by Eq.~\eqref{eq:def-torque}.
Therefore, the wall can be regarded as a boundary between two domains with distinct topological invariants.
In other words, the torque term in Eq.~\eqref{eq:def-torque}, with an appropriately chosen $\alpha$, renders the potential walls topological domain walls.
This explains why the system exhibits edge currents that are robust against changes in wall geometry and the introduction of defects for certain values of $\alpha$ and $\Omega$.

The edge currents observed in the phase-separated states of interacting particles, discussed in Sec.~\ref{sec:collective_behavior}, can also be understood in the same way.
We define the effective torque $\Omega'_j = \Omega + T_j$, which corresponds to $\Omega(\rho)$ in the hydrodynamic description.
In Fig.~\ref{fig:torque_sgn}(c), we show the spatial distribution of the coarse-grained effective torque $\Omega'$ for a phase-separated state, obtained by averaging $\Omega'_j$ over time and space.
We also show in Fig.~\ref{fig:torque_sgn}(d) a histogram of $\Omega'_j$ inside and outside the droplet in the same phase-separated state.
Inside the droplet, the effective chirality $\Omega'$ is predominantly negative, whereas outside the droplet it is positive.
Since the Chern numbers are given by $\mathcal C_\pm = \mp 2\RM{sgn}[\Omega(\rho_0)] \simeq \mp 2\RM{sgn}[\Omega']$, this phase separation can be regarded as the coexistence of two topologically distinct phases.
Hence, we conclude that the edge currents observed in the phase-separated state are also topological modes.
We remark on the difference between phase separation in our model and MIPS in chiral active Brownian particles reported in Refs.~\cite{Ma2022JCP, Metzger2026}.
Even in MIPS with chirality, ``apparent'' edge currents have been observed~\cite{Ma2022JCP, Metzger2026}; however, they are not topologically protected because the chirality is the same in both the dense and dilute phases, i.e., the two phases are topologically equivalent.
We also note that the same topological interpretation cannot be directly applied to the clustering states shown in Figs.~\ref{fig:config}(c) and (d), which are characterized by a finite hexatic order parameter.
This is because such hexatically ordered phases are beyond the scope of the hydrodynamic equations.
However, in Appendix~\ref{ape_ssec:omega_sgn_cluster}, we show that such hexatic cluster states also exhibit a similar separation into two regimes with opposite chirality.
This implies that edge modes may also exist in the clustering states.

\section{Conclusions}
\label{sec:summary}
In this work, we have introduced a simple particle model of two-dimensional chiral active swimmers that switch their chirality through feedback from interactions.
We have studied the behavior of a single particle under confinement and collective behavior in bulk.
For the single particle case, we have demonstrated that owing to chirality switching, the model shows edge currents that are immune to changes in wall geometries and the presence of defects. 
The edge localization and currents can be understood from the simple diffusion equation that contains odd diffusivity and mobility. 
For the interacting particles, we have found that chirality switching leads to phase separation accompanied by edge currents along the interface.
We have then constructed an effective hydrodynamic theory and qualitatively explained phase separation induced by chirality switching and the existence of the edge currents.
We have further uncovered that for both single- and many-body cases, the robustness of the edge currents can be understood using the language of topological band theory.
By regarding the linearized hydrodynamic equations as a Schr\"odinger-like equation with an effective Hamiltonian, one can calculate the Chern numbers as topological invariants.
In our system, the topological invariant depends solely on the sign of the effective chirality, which includes the effects of interactions.
We have shown that the robustness of the edge currents for a single particle originates from the fact that potential walls act as topological domain walls separating regions with opposite Chern numbers.
Furthermore, we have demonstrated that phase separation observed in interacting particles is a novel phenomenon that can be understood as the coexistence of two topologically distinct domains.
In hindsight, since the topological invariant is determined solely by the sign of the chirality, chirality switching provides a natural mechanism for creating topological domain walls that support edge currents.

Edge currents in chiral active matter have been studied since the early stages of the field.
More recently, phase separation accompanied by edge currents has also been actively investigated~\cite{Caporusso2024PRL, Adorjani2024EPJ, Siebers2024PNAS, Zhou2025PNASNex, Caprini2025JCP, Metzger2026, wang2026edge}.
In parallel, topological edge states in active matter have attracted increasing attention~\cite{Sone_2026}, and recent studies have reported their realization in lattice models~\cite{Osat2026}.
This study has established a route to constructing topological edge states in a particle-based model.
In particular, the spontaneous separation of interacting particles into two topologically distinct phases opens a new avenue connecting three major themes in active matter: phase separation, edge currents, and topology.
These results also suggest several important directions for future research.
One immediate question concerns experimental systems that exhibit phase separation accompanied by edge currents, such as magnetic colloids~\cite{Soni2019NatPhys, Massana-Cid2021PRR, Katuri2024CommPhys}.
It would be interesting to examine whether these phase-separated states can also be understood as the coexistence of two topologically distinct domains.
Another direction is to develop a scalar field theory for such topological phase separation.
Our hydrodynamic equation can be reduced to a scalar field theory with odd terms, but topology cannot be discussed in the same way as in the linearized equations because all odd terms are nonlinear.
This may motivate future studies of topology in chiral active fluids based on nonlinear eigenvalue problems~\cite{Sone2024NatPhys}.
A further important direction is to seek ways to experimentally design chirality switching in order to realize topologically distinct domains.
Here we have already demonstrated that an anisotropic chiral walker effectively switches chirality when colliding with a wall (Appendix~\ref{ape:experiment}), opening new strategies for exploiting topological modes in collectives of motile microrobots.

\begin{acknowledgments}
The authors acknowledge support by the state of Baden-Württemberg through bwHPC
and the German Research Foundation (DFG) through grant INST 35/1597-1 FUGG.
In addition, Y.K. acknowledges support from the Japan Society for the Promotion of Science (JSPS) through the Overseas Research Fellowships. 
E.M. and T.S. acknowledge funding through the Deutsche Forschungsgemeinschaft (DFG, German Research Foundation) within the framework of the collaborative research centers “Multiscale Simulation Methods for Soft-Matter Systems” (TRR 146) under Project No.233630050. E.M. thanks the International Max-Planck Research School for Intelligent Systems (IMPRS-IS) for support. 
\end{acknowledgments}

\appendix
\section{Effects of rotational diffusion on edge currents of a single particle}
\label{ape:effects_of_Dr}
\begin{figure}[b!]
\centering
\includegraphics[width=1.0\linewidth]{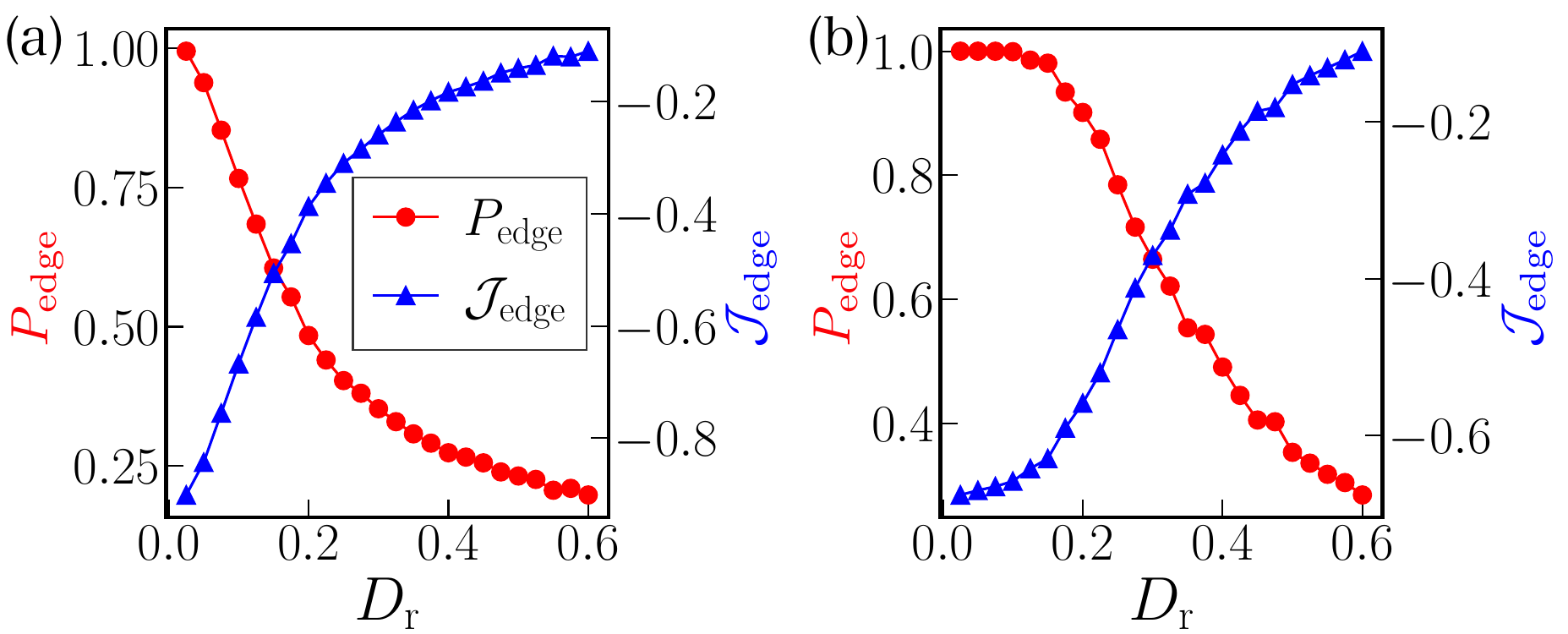}
\caption{\label{fig:single_Dr} Edge occupation probability and total edge current as functions of $D_\RM{r}$ at (a)~$\Omega = 0.25$ and (b)~$\Omega = 1$, with $\alpha = 2$.
The simulations are performed in the circular confinement shown in Figs.~\ref{fig:histogram}(a) and (d) (the same setup as in Fig.~\ref{fig:PD_single}).
}
\end{figure}
In Fig.~\ref{fig:PD_single} in the main text, we show the edge occupation probability and the total edge current for a fixed rotational diffusion constant, $D_\RM{r}=0.1$.
Figure~\ref{fig:single_Dr} shows $P_\RM{edge}$ and $\mathcal J_\RM{edge}$ as functions of $D_\RM{r}$ at fixed $\alpha$ and $\Omega$.
As $D_\RM{r}$ increases, both edge localization and the edge current disappear.
This is natural because the model with $T_j(t)=0$ approaches an equilibrium system, namely a passive Brownian particle, in the limit of large $D_\RM{r}$.
The transition from a localized to a delocalized state upon varying $D_\RM{r}$ is gradual.

\section{Experiment}
\label{ape:experiment}
\begin{figure}[t!]
\centering
\includegraphics[width=1.0\linewidth]{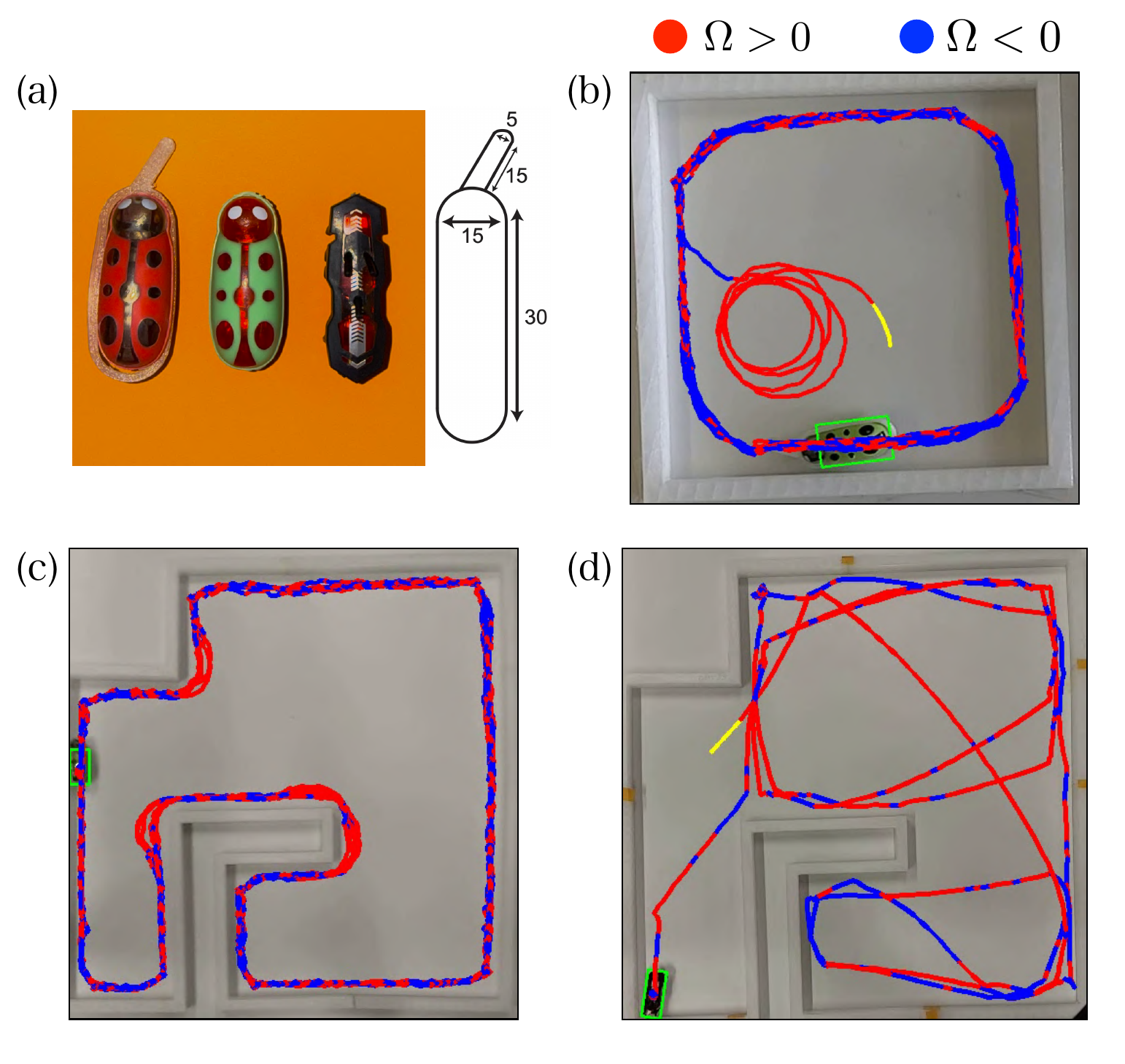}
\caption{\label{fig:experiment}Experimental demonstration of an edge current induced by chirality switching.
(a)~Walkers used for experiments (see Appendix~\ref{ape:experiment} for details of the experimental setup).
(b,c)~Typical trajectories of a chiral walker recorded within (b)~square and (c)~complex confinements.
(d)~Typical trajectory of an achiral walker recorded within the complex confinement.
In panels (b)-(d), each data point is colored according to the sign of the chirality.
For the experiment shown in panel~(b), the attachment is installed on the walker. 
Note that the attachment on the walker increases the probability that the walker reaches the edge from the bulk; otherwise, it does not contribute to the formation of the edge current.
}
\end{figure}
In this appendix, we demonstrate that chirality switching indeed leads to an edge current using a commercially available beetle toy as a chiral walker.
A single walker measures $45~\RM{mm} \times 15~\RM{mm} \times 15~\RM{mm}$. 
To enhance its interaction with the boundary walls, a custom 3D-printed enclosure [Fig.~\ref{fig:experiment}(a)] was mounted on top of the walker as a hat-like attachment. 
Note that the hat does not play a role in developing an edge current; rather, it increases the probability of coupling of the walker from the bulk to the edge.
Locomotion is generated by a vibration motor embedded within the walker body. The motor induces vertical vibrations, which, in combination with the walker's asymmetric legs, produce forward propulsion when the walker is placed on a rigid surface such as a wooden desk. 
The resulting motion is inherently chiral and this chirality is a feature of the commercially available design. 
While the radius and sense of rotation can be adjusted by modifying the lengths of the legs on one side of the walker, all experiments reported here were performed using the unmodified walker.
Experiments were carried out on a wooden desk within a square enclosure measuring $200~\RM{mm}\times 200~\RM{mm} \times 15~\RM{mm}$ ($400~\RM{mm} \times 400~\RM{mm} \times 15~\RM{mm}$ for the arena with a complex boundary).
The enclosure was fabricated using a Prusa i3 MK2S 3D printer with $1.75~\RM{mm}$ diameter polylactic acid (PLA) filament and a layer height of $0.2~\RM{mm}$. 
Videos were recorded at 30 frames per second using an overhead camera.
The walker's position and orientation were extracted from the recorded videos using custom Python scripts based on OpenCV and the connected-components functionality available in scikit-image. 
The resulting detections were linked into trajectories using TrackPy, followed by smoothing to reduce localization noise.
The sign of the chirality was obtained by considering three relatively close points along the trajectory, $\RM{sgn}[\Omega] = \RM{sgn} [(\dd\bm{x}_1\times \dd\bm{x}_2)\cdot\bm{e_z}]$, where $\dd\bm{x}_1 = \bm{x}_{t_2}-\bm{x}_{t_1}$ and $\dd\bm{x}_2 = \bm{x}_{t_3}-\bm{x}_{t_2}$, for $t_1 < t_2 <t_3$, and $\bm{x}=(x,y)$ is the position of the walker. 

In Fig.~\ref{fig:experiment}(b), we show a typical trajectory of a walker within a square confinement, extracted from a recorded video.
The color indicates the sign of the chirality. 
When the walker is away from the wall, it undergoes counterclockwise motion.
Upon contacting the wall, it switches its chirality and begins to move along the confining wall (see Video~S6).
The sign of the chirality alternates between positive and negative as the walker moves along the wall.
We further show in Fig.~\ref{fig:experiment}(c) the trajectory of a chiral walker within a complex confinement (see Video~S7).
For comparison, Fig.~\ref{fig:experiment}(d) shows the trajectory of an achiral walker within the same confinement (see Video~S8).
The chiral walker clearly exhibits an edge current, demonstrating the robustness of edge currents induced by chirality switching against the confinement geometry.
These trajectories resemble those observed in the particle-model simulations, such as the trajectory shown in Fig.~\ref{fig:model}(b) of the main text.
Of course, our particle model is not intended to provide an exact description of this experimental system.
Nevertheless, this experiment suggests that chirality switching plays a key role in the emergence of robust edge currents.

\section{Effective diffusion equation for a single particle under confinement}
\label{ape:current_single}
Here, we drive Eq.~\eqref{eq:current_single} in the main text.
We consider noninteracting particles under confinement, modeled by the wall force $\bm F_\RM{wall}(\bm r) = -\nabla U_\RM{wall}(\bm r)$.
The Smoluchowski equation corresponding to Eqs.~\eqref{eom1} and \eqref{eom2} is given by
\begin{equation}
\begin{split}
\p_t \Psi =  &-\nabla\cdot \qty[
\mu \bm F_\RM{wall}(\bm r)   + v_0\bm e(\phi)] \Psi \\
&- \p_\phi  \qty[\Omega + \alpha \bm F_\RM{wall}(\bm r)\cdot \bm e(\phi) -D_\RM{r}  \p_\phi]\Psi,
\end{split}
\end{equation}
where $\Psi = \Psi(\bm r,\phi,t)$ is distribution function defined by $ \Psi= \expval{\delta(\bm r-\bm r(t))\delta_{2\pi}(\phi-\phi(t))}$. 
Here, $\delta_{2\pi}(\phi) = \sum_{n\in\mathbb Z}\delta(\phi + 2\pi n)$. 
By integrating the Smoluchowski equation over $\phi$, we have a continuity equation for the spatial distribution $\rho(\bm r,t) = \int_{0}^{2\pi}\dd\phi\ \Psi(\bm r,\phi,t)$:
\begin{equation}
\p_t\rho(\bm r,t) = -\nabla\cdot \bm J(\bm r,t)
\end{equation}
with the current
\begin{equation}
\bm J(\bm r,t) = \mu \bm F_\RM{wall}(\bm r)\rho(\bm r,t) + v_0\bm p(\bm r,t), \label{eq:current_single2}
\end{equation}
where we define the polarization as $\bm p(\bm r,t) = \int_{0}^{2\pi}\dd\phi\ \bm e(\phi)\Psi(\bm r,\phi,t)$. 
From the definition of $\bm p$ and the Smoluchowski equation, its dynamics reads
\begin{equation}
\begin{split}
\p_t \bm p = - D_\RM{r}\bm p -v_0 \nabla\cdot \qty(\mathsf Q +\frac{\rho}{2}\mathbbm 1)
- \mu \nabla\cdot\qty(\bm F_\RM{wall} \bm p)
\\
- \Omega \bm p^\perp - \alpha\underline{\bm\epsilon}\qty(\mathsf Q +\frac{\rho}{2}\mathbbm 1)\bm F_\RM{wall}, \label{eq:polarization_single1}
\end{split}
\end{equation}
where $\mathbbm 1$ is the identity matrix, $\underline{\bm\epsilon}$ represents the Levi--Civita tensor whose elements are given by $\epsilon_{x,y} = - \epsilon_{y,x}=1$ and  $\epsilon_{x,x} = \epsilon_{y,y}=0$, and $\bm p^\perp = \underline{\bm\epsilon}\bm p = (p_y,-p_x)$.
The tensor $\mathsf Q$ is the second moment of the orientation defined by $\mathsf Q(\bm r,t) = \int_{0}^{2\pi}\dd\phi\ \qty[\bm e(\phi)\bm e(\phi)- \mathbbm 1/2]\Psi(\bm r,\phi,t)$. 
To obtain a closed equation for $\rho(\bm r,t)$, we assume that $\mathsf Q\simeq \mathsf O$, since $\mathsf Q$ relaxes faster than $\bm p$.
To eliminate $\bm p$, we use the adiabatic approximation, $\p_t\bm p\simeq 0$.
Equation~\eqref{eq:polarization_single1} is then approximated as
\begin{equation}
\bm p\simeq -\frac{v_0}{2}\mathsf W^{-1}\nabla \rho - \frac{\alpha}{2}\rho\mathsf W^{-1}\bm F_\RM{wall}^\perp - \mu \mathsf W^{-1}\nabla\cdot\qty(\bm F_\RM{wall} \bm p), \label{eq:polarization_single2}
\end{equation}
where
\begin{equation}
\mathsf W = \mqty(D_\RM{r} & \Omega \\ -\Omega & D_\RM{r})
\end{equation}
and $\bm F_\RM{wall}^\perp = \underline{\bm\epsilon}\bm F_\RM{wall}$.
Since $\bm F_\RM{wall} = -\nabla U_\RM{wall}$, the first two terms in Eq.~\eqref{eq:polarization_single2} are $O(\nabla)$, whereas the third term is $O(\nabla^3)$. 
Therefore, using Eqs.~\eqref{eq:current_single2} and \eqref{eq:polarization_single2}, we obtain the density current given in Eq.~\eqref{eq:current_single}, approximated up to $O(\nabla)$.
\section{Many-body simulations}
\label{ape:SM_data}
\begin{figure}[b!]
\centering
\includegraphics[width=8.5cm]{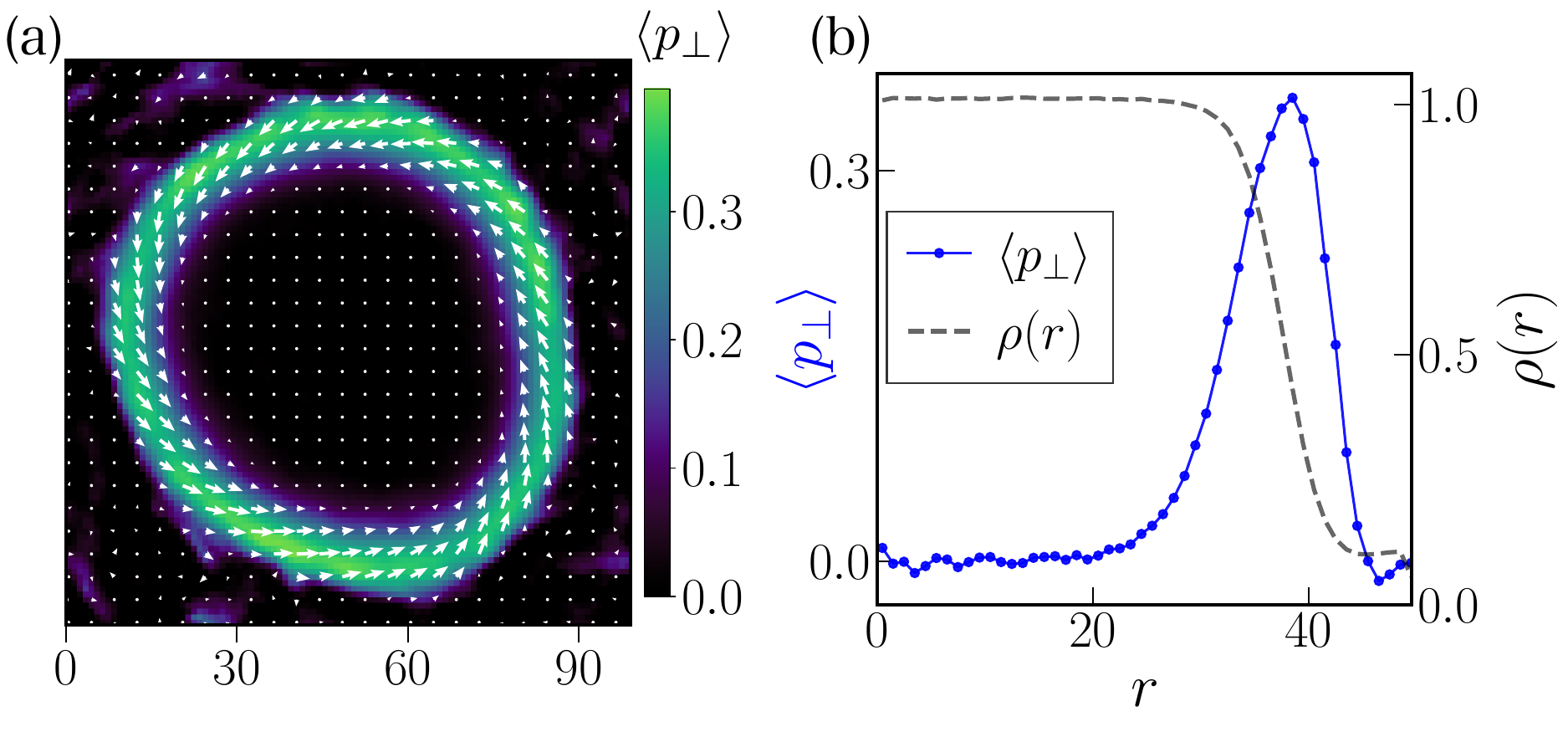}
\caption{\label{fig:polarization} Local polarization in a phase-separated state.
(a)~Color map of the spatially averaged tangential component of the polarization.
The arrows represent the average local polarization.
(b)~Average tangential polarization $\expval{p_\perp}$ as a function of the distance from the center of the droplet.
The parameters are set to $\alpha=6$, $\Omega=2$, $\varphi=0.4$, $D_\RM{r}=0.1$, and $N=5\times 10^3$ (the same as those in Fig.~\ref{fig:collective_current}).
}
\end{figure}
\begin{figure}[t]
\centering
\includegraphics[width=8.5cm]{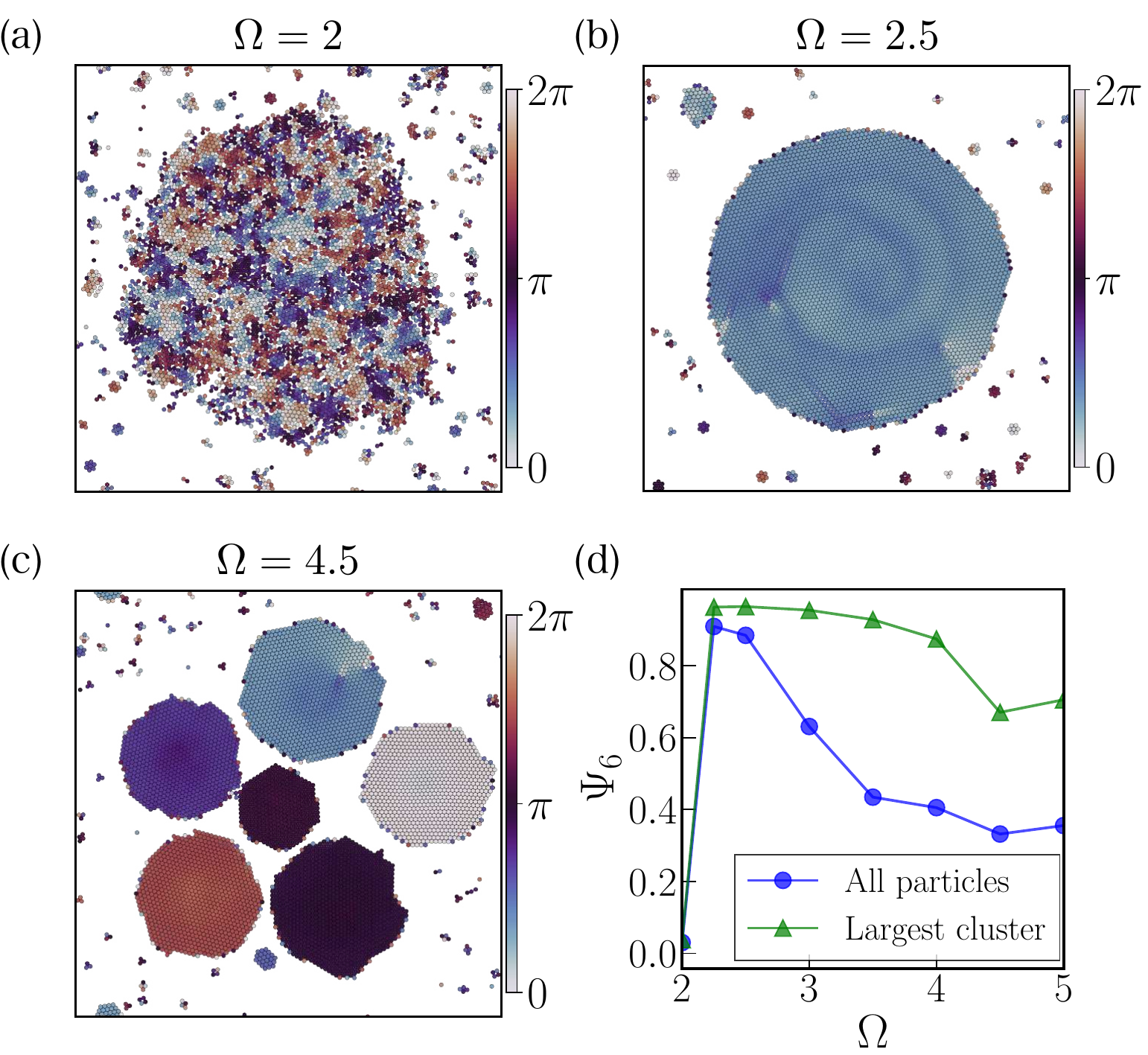}
\caption{\label{fig:hexatic}Particle configurations colored according to the phase of the local hexatic order parameter, $\theta_6(\bm r_j) = \RM{arg}[\psi_6(\bm r_j)]$, at (a)~$\Omega=2$, (b)~$\Omega=2.5$, and (c)~$\Omega=4.5$.
(d)~Comparison of the hexatic order parameters averaged over all particles and over the particles belonging to the largest cluster. The parameters are set to $\alpha=6$, $\varphi=0.4$, $D_\RM{r}=0.1$, and $N=5\times 10^3$ [the same as those in Fig.~\ref{fig:config}(g)].
}
\end{figure}
\begin{figure}[t]
\centering
\includegraphics[width=8.5cm]{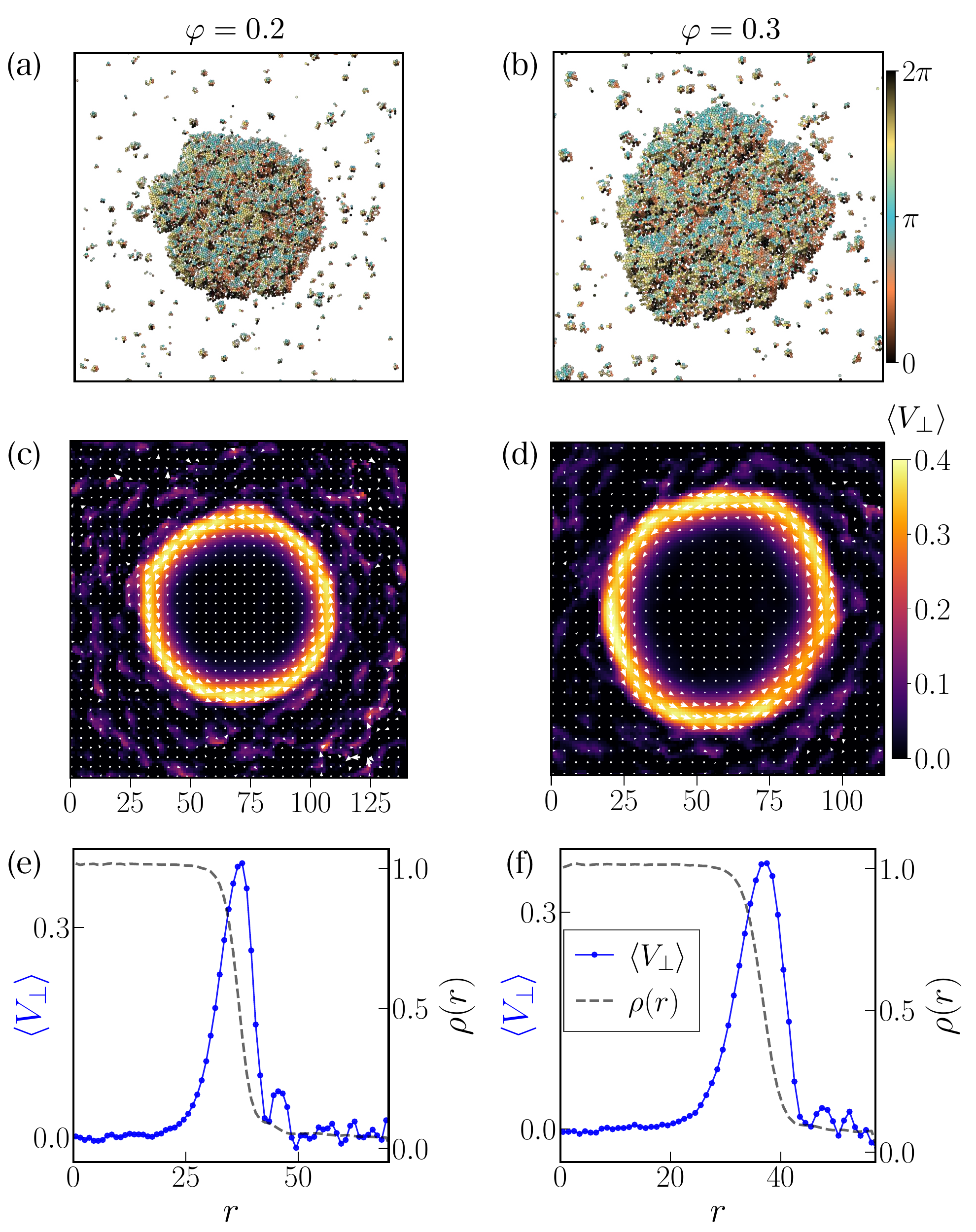}
\caption{\label{fig:density_dependence} Density dependence of phase separation induced by chirality switching.
(a,b)~Typical particle configurations at $\varphi=0.2$ and $0.3$, respectively.
The color indicates the polar angle of the velocity $\dot{\bm r}_j$.
(c,d)~Spatial profiles of the average tangential velocity at $\varphi=0.2$ and $0.3$, respectively.
(e,f)~Average tangential velocity as a function of the distance from the center of the droplet at $\varphi=0.2$ and $0.3$, respectively.
The dashed lines represent the average density profiles.
The other parameters are the same as those in Fig.~\ref{fig:config}(b) and Fig.~\ref{fig:collective_current}: $\alpha=6$, $\Omega=2$, $D_\RM{r}=0.1$, and $N=5\times 10^3$.
}
\end{figure}
\begin{figure}[t]
\centering
\includegraphics[width=8.5cm]{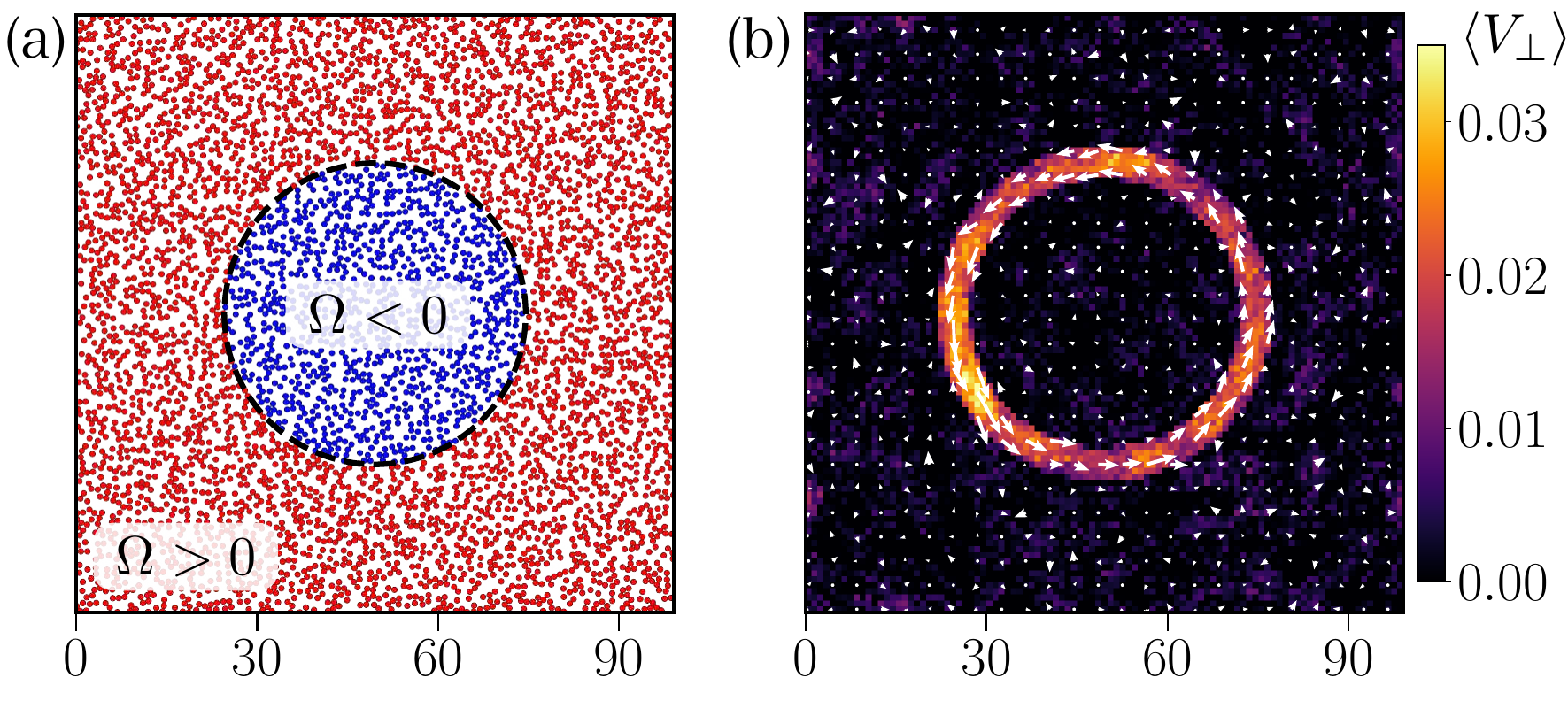}
\caption{\label{fig:domain} Interacting chiral active Brownian particles with a position-dependent torque $\Omega(\bm r_j)$.
We define two domains separated by an interface, with $\Omega(\bm r_j)<0$ in one domain and $\Omega(\bm r_j)>0$ in the other.
The dynamics are governed by Eqs.~\eqref{eom1} and \eqref{eom2} without $T_j$.
(a)~Typical particle configuration for the case in which the domain wall is a circle of radius $25\sigma$ centered at $(L_x/2,L_y/2)$.
The particles with $\Omega>0$ and $\Omega<0$ are shown in red and blue, respectively.
The dashed circle represents the domain boundary.
(b)~Average tangential velocity $\expval{V_\perp}$.
The arrows represent the average local velocity.
The parameters are set to $|\Omega|=2$, $\varphi=0.4$, $D_\RM{r}=0.1$, and $N=5\times 10^3$.
}
\end{figure}
\begin{figure}[t]
\centering
\includegraphics[width=1.0\linewidth]{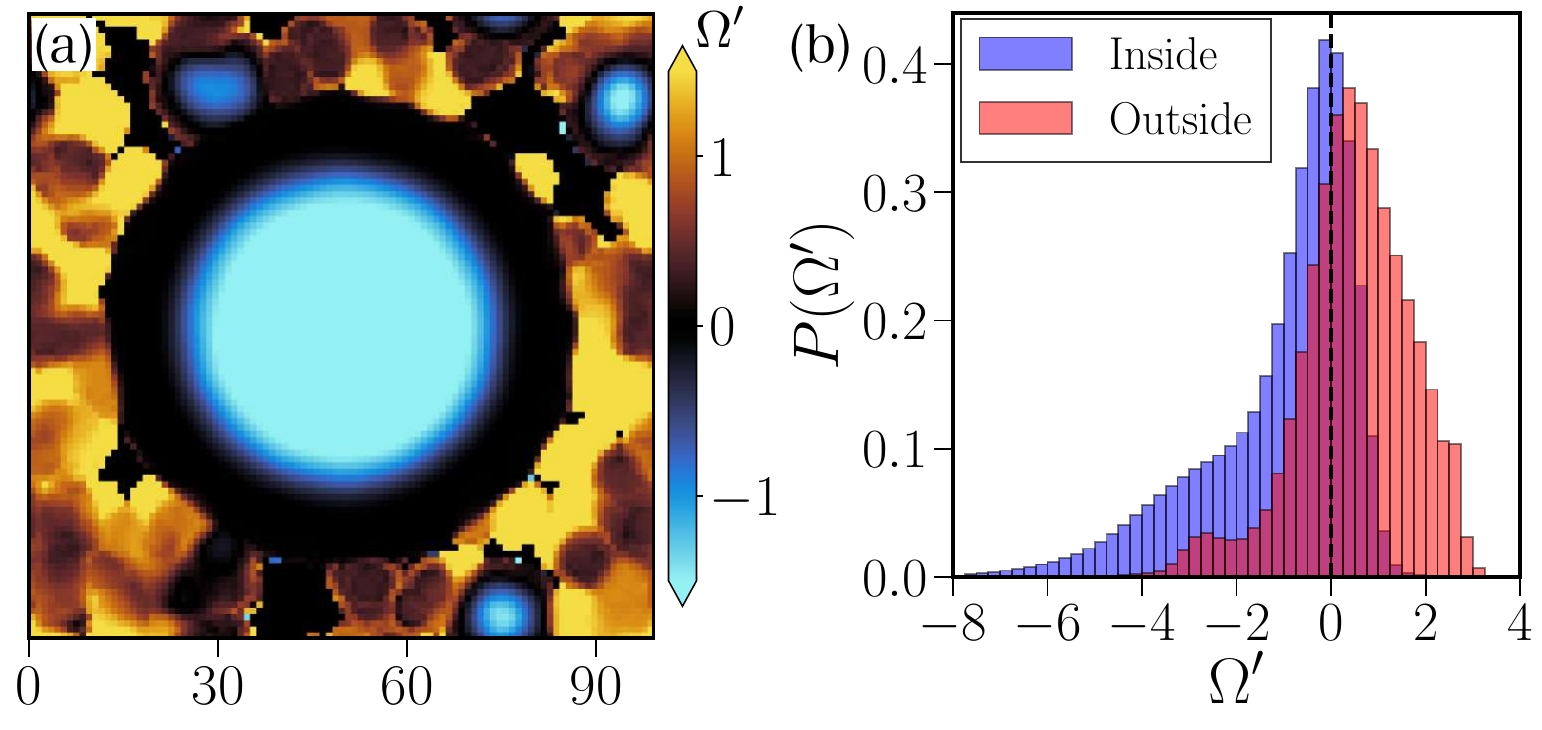}
\caption{\label{fig:torque_sgn_cluster}
Local effective torque $\Omega'$ in a hexatically structured clustering state.
(a)~Spatial profile of the average effective torque $\Omega'$.
(b)~Histogram of the local effective torque $\Omega'_j = \Omega + T_j$ inside and outside dense clusters.
The parameters are set to $\alpha=6$, $\Omega=2.5$, $\varphi=0.4$, $D_\RM{r}=0.1$, and $N=5\times 10^3$ [the same as those in Fig.~\ref{fig:config}(c)].
}
\end{figure}
\subsection{Local polarization in a phase-separated state}
\label{ape_ssec:polarization}
Figure~\ref{fig:polarization} depicts the local polarization in the phase-separated state shown in Fig.~\ref{fig:collective_current}.
We calculate the local polarization by taking the spatial average of $\bm e(\phi_j)$ over a small circular region of radius $3\sigma$.
We see that the behavior of the local polarization is the same as that of the velocity field shown in Fig.~\ref{fig:collective_current}.
Thus, the edge currents can also be discussed in terms of the local polarization.
We also remark that this behavior is distinct from that in MIPS, where the local polarization points toward the inside of the droplet~\cite{Richard2016SoftMat, Speck2020SoftMat}.

\subsection{Hexatic order parameter}
\label{ape_ssec:hexatic}
In Fig.~\ref{fig:config}(g) in the main text, we show the global hexatic order parameter averaged over all particles.
As $\Omega$ increases, $\Psi_6$ jumps to a finite value and subsequently decreases.
The decrease in $\Psi_6$ at large $\Omega$ stems from the presence of multiple dense clusters.
To illustrate this, Figs.~\ref{fig:hexatic}(a)-(c) present particle configurations at three different values of $\Omega$, colored according to the phase of the local hexatic order parameter, $\theta_6(\bm r_j) = \RM{arg}[\psi_6(\bm r_j)]$.
In the phase-separated state, no hexatic order is observed [Fig.~\ref{fig:hexatic}(a)], as discussed in the main text.
In the rotating cluster states, each cluster exhibits hexatic order.
When multiple clusters are present, the hexatic order parameters of the individual clusters have different phases [see Fig.~\ref{fig:hexatic}(c)].
Therefore, when $\Psi_6$ is calculated over all particles in such multiple-cluster states, contributions from different clusters may cancel due to their phase differences.
Figure~\ref{fig:hexatic}(d) shows $\Psi_6$ calculated by averaging over all particles and over particles belonging to the largest cluster.
The hexatic order of the largest cluster depends only weakly on $\Omega$.

\subsection{Density dependence of phase separation with edge currents}
\label{ape_ssec:density_dependence}
In Sec.~\ref{sec:collective_behavior}, we primarily discuss phase separation at the packing fraction $\varphi=0.4$.
In Fig.~\ref{fig:density_dependence}, we show that the phenomenology discussed in Sec.~\ref{sec:collective_behavior} is insensitive to changes in the packing fraction.
Figures~\ref{fig:density_dependence}(a) and (b) show typical particle configurations at $\varphi=0.2$ and $0.3$, respectively.
In both cases, we observe phase separation accompanied by edge currents, as shown in Figs.~\ref{fig:density_dependence}(c)-(f).
As in standard liquid-vapor phase separation, varying the overall density does not change the densities of the liquid and gas phases, but only their relative fractions [see the density profiles shown in Figs.~\ref{fig:density_dependence}(e) and (f)].

\subsection{Current along a domain wall in interacting particles}
\label{ape_ssec:domain}
In Fig.~\ref{fig:torque_sgn}(b) in the main text, we show that a single chiral active Brownian particle exhibits edge currents along the domain boundary in a system composed of two domains with opposite signs of chirality.
The same behavior is observed even for interacting particles.
To demonstrate this, we simulate interacting particles governed by Eqs.~\eqref{eom1} and \eqref{eom2} without the torque term $T_j$.
The setup is the same as that in Sec.~\ref{sec:collective_behavior}.
We set two domains with opposite signs of chirality, as shown in Fig.~\ref{fig:domain}(a).
Figure~\ref{fig:domain}(b) shows the spatial profile of the velocity component perpendicular to the vector from the center of the system.
One observes that the tangential velocity is finite only near the domain wall and is close to zero elsewhere.

\subsection{Local effective chiral torque in a clustering state}
\label{ape_ssec:omega_sgn_cluster}
In Figs.~\ref{fig:torque_sgn}(c) and (d) in the main text, we show the local effective chirality $\Omega'$ in a phase-separated state, where the inside of the droplet exhibits a liquid-like structure.
As discussed in Sec.~\ref{sec:collective_behavior}, the system also exhibits clustering states, in which multiple dense clusters coexist and their interiors exhibit a hexatic crystalline-like structure.
Figure~\ref{fig:torque_sgn_cluster}(a) shows the local effective chirality, obtained by taking the spatiotemporal average of $\Omega'_j = \Omega + T_j$, in such a clustering state, corresponding to the state shown in Fig.~\ref{fig:config}(c).
As in the phase-separated state, $\Omega'$ is negative inside the dense clusters, whereas it is positive outside.
In Fig.~\ref{fig:torque_sgn_cluster}(b), we also show the histogram of the effective torque of each particle, $\Omega'_j$, inside and outside the dense clusters.
Although the spatiotemporally averaged $\Omega'$ clearly shows separation into two domains with negative and positive $\Omega'$, the distribution of $\Omega'_j$ is less clearly separated than that in the phase-separated state.
These observations suggest the possible existence of topological edge modes even in clustering states.
However, we note that the argument for topological edge modes in Sec.~\ref{sec:topo} cannot be applied directly, since the hexatic structure is beyond the scope of the hydrodynamic equations.

\section{Hydrodynamic equations for interacting particles}
\label{ape:hydro}
\begin{figure*}[t]
\centering
\includegraphics[width=15.5cm]{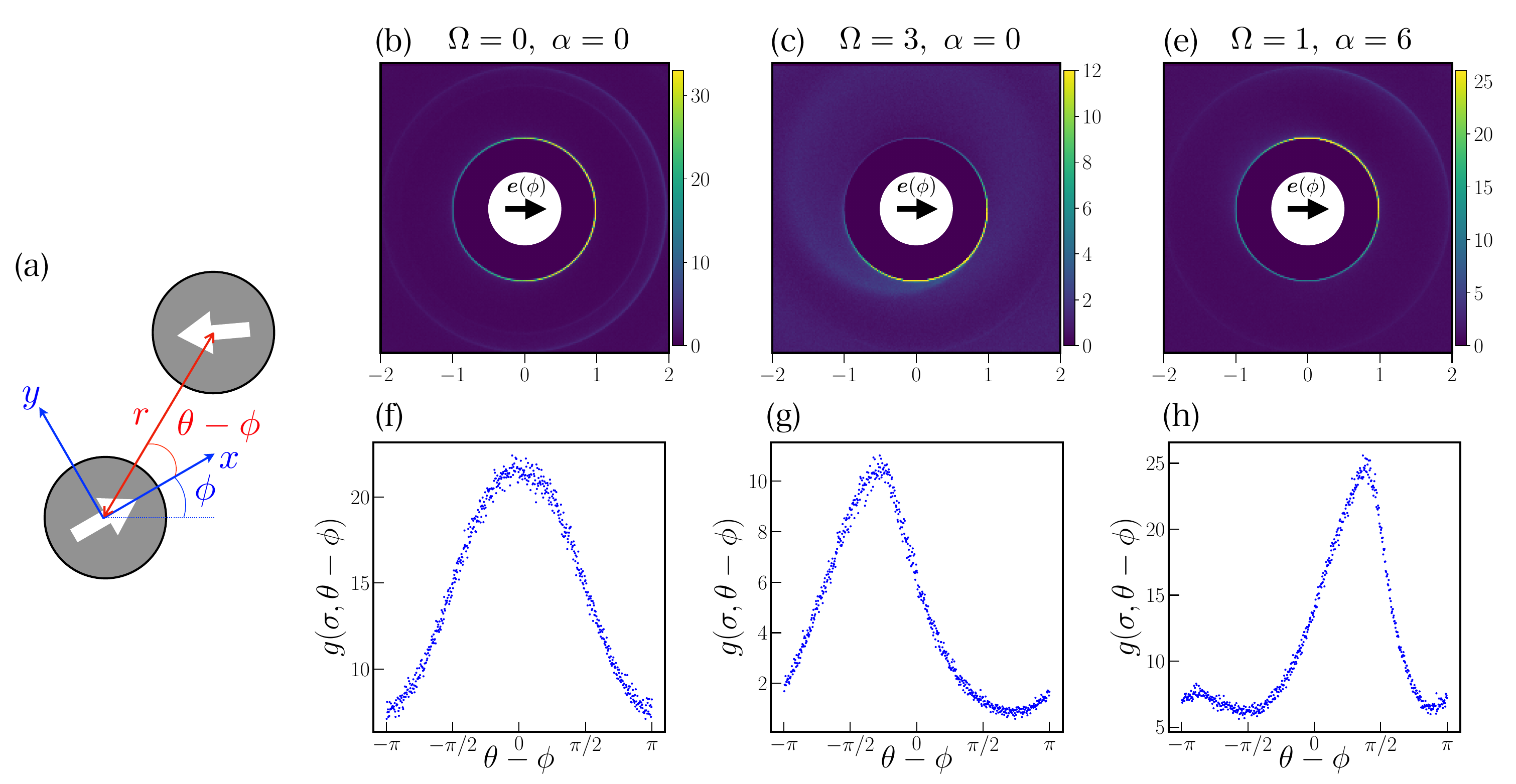}
\caption{\label{fig:gr}
(a)~Schematic illustration of the local frame used to compute the pair distribution function.
The pair distribution function $g(\bm r,\phi)$ in homogeneous states, measured in the coordinate system of the tagged particle with the $x$-axis taken parallel to its orientation, is shown for
(b)~$\Omega=0,\ \alpha=0$, (c)~$\Omega=3,\ \alpha=0$, and (d)~$\Omega=1,\ \alpha=6$.
The white circle represents the tagged particle.
The bottom row shows $g(r,\theta,\phi)$ at $r=\sigma$, where $\theta$ is the polar angle of $\bm r$:
(f)~$\Omega=0,\ \alpha=0$, (g)~$\Omega=3,\ \alpha=0$, and (h)~$\Omega=1,\ \alpha=6$.
The other parameters are set to $D_\RM{r}=0.1$, $\varphi=0.4$, and $N=5\times 10^3$.
}
\end{figure*}
\subsection{Closure approximation}
\label{ape_ssec:closure}
In this section, we derive Eq.~\eqref{eq:one-body-2} in the main text.
Here, we roughly follow Refs.~\cite{Bialk2013EPL, Speck2015JCP}.
The one-body distribution function is defined by
\begin{equation}
\Psi(\bm r,\phi,t) = \expval{\sum_{j=1}^{N} \delta(\bm r-\bm r_j(t)) \delta_{2\pi}(\phi- \phi_j(t))},
\end{equation}
where $\delta_{2\pi}(\phi) = \sum_{n\in\mathbb Z}\delta(\phi + 2n\pi)$.
Using the It\^o formula and Eqs.~\eqref{eom1} and \eqref{eom2}, the time derivative of $\Psi(\bm r,\phi,t)$ reads
\begin{equation}
\begin{split}
\p_t \Psi =  &-\nabla\cdot \qty[
\mu \bm G  + v_0\bm e(\phi)] \Psi \\
&- \p_\phi  \qty[\Omega + \alpha \bm G\cdot \bm e(\phi) -D_\RM{r}  \p_\phi]\Psi. \label{eq:one-body-1}  
\end{split}
\end{equation}
Here, $\bm G = \bm G(\bm r,\phi,t)$ is so-called conditional force defined by~\cite{Bialk2013EPL, Speck2020SoftMat}
\begin{align}
\bm G(\bm r,\phi,t) &= \int_{V}\dd[2]\bm r'\int_{0}^{2\pi}\dd\phi'\ \bm f(\bm r-\bm r')\Psi_2(\bm r', \phi' | \bm r, \phi;t) \notag \\
&= \int_{V}\dd[2]\bm r'\ \bm f(\bm r-\bm r')\Psi_2(\bm r' | \bm r, \phi;t), \label{eq:conditional-force}
\end{align}
where $\bm f(\bm r) = -\nabla U(r)$, 
$\Psi_2(\bm r', \phi' | \bm r, \phi; t) = \Psi_2(\bm r, \phi; \bm r',\phi' ;t) /\Psi(\bm r,\phi;t) $ is the conditional two-body distribution function, and $\Psi_2(\bm r', \phi';\bm r, \phi; t)$ is the two-body distribution function. 
The second equality of Eq.~\eqref{eq:conditional-force} follows from $\Psi_2(\bm r' | \bm r, \phi;t) = \int_{0}^{2\pi}\dd\phi'\ \Psi_2(\bm r', \phi' | \bm r, \phi; t)$.
Since Eq.~\eqref{eq:conditional-force} contains $\Psi_2(\bm r' | \bm r, \phi;t)$, Eq.~\eqref{eq:one-body-1} is not closed for $\Psi(\bm r,\phi,t)$. 
Here, we truncate the BBGKY-type hierarchy following Refs~\cite{Bialk2013EPL}.
We first define the pair distribution function $g(\bm r',\bm r, \phi;t)$ through the following equation~\cite{Hansen}:
\begin{equation}
\Psi_2(\bm r' | \bm r,\phi;t)= g(\bm r' , \bm r, \phi;t)\rho(\bm r',t),
\end{equation}
where $\rho(\bm r)$ is the density field defined by $\rho(\bm r,t) = \expval*{\sum_{j=1}^{N}\delta(\bm r-\bm r_j(t))}$.
We assume the time dependence in the pair distribution function to be negligible and the homogeneous condition to be satisfied: $g(\bm r', \bm r, \phi;t)\simeq g(\bm r'-\bm r, \phi)$.
Equation~\eqref{eq:conditional-force} then becomes
\begin{align}
\bm G(\bm r,\phi,t) &=  \int_{V}\dd[2]\bm r'\ \bm f(\bm r-\bm r')g(\bm r'-\bm r, \phi)\rho(\bm r',t) \notag \\
&= - \int_{r < r_\RM{c}}\dd[2]\bm r'\ \bm f(\bm r')g(\bm r', \phi)\rho(\bm r + \bm r',t).
\end{align}
In the second equality, we used $\bm f(\bm r) = - \bm f(-\bm r)$ and assumed that the interaction is short-ranged, such that $U(r) = 0$ for $r>r_\RM{c}$, where $r_\RM{c}$ is sufficiently small compared with the system size.
Since $r_\RM{c}$ is small, one can expand the density field in the integral as
\begin{equation}
\rho(\bm r + \bm r') = \rho(\bm r) + \bm r'\cdot \nabla\rho(\bm r) + O(\nabla ^2). \label{eq:gradient_expansion}
\end{equation}
Neglecting the higher gradient terms $O(\nabla^2)$, we have
\begin{equation}
 G_\alpha(\bm r,\phi,t) \simeq  g^{(0)}_\alpha(\phi)\rho(\bm r,t) - g_{\alpha,\beta}^{(1)}(\phi)\p_\beta\rho(\bm r,t)
\end{equation}
with 
\begin{align}
g^{(0)}_\alpha(\phi)&= - \int_{r < r_\RM{c}}\dd[2]\bm r\  g(\bm r, \phi)f_\alpha(\bm r),\label{eq:averaged-force1} \\ 
g_{\alpha,\beta}^{(1)}(\phi) &=  \int_{r < r_\RM{c}}\dd[2]\bm r\  g(\bm r, \phi)f_\alpha (\bm r) x_\beta . \label{eq:averaged-force2}
\end{align} 
In the case of active Brownian particles without chirality, the pair distribution function $g(\bm r, \phi)$ is axisymmetric with respect to the orientation $\bm e(\phi)$ and takes larger values in the forward direction~\cite{Bialk2013EPL, Poncet2021PRE}.
This suggests that $g(\bm r,\phi) \simeq \tilde g(r,\theta - \phi)$ and $\tilde g(\cdot, \phi) =\tilde g(\cdot, -\phi)$, where $\theta$ is the polar angle of $\bm r$.
This assumption has also been employed for chiral active particles~\cite{Ma2022JCP, Sansa2022JCP}, and we adopt it in the following.
However, this assumption should be modified when $\Omega$ or $\alpha$ is large.
In Appendix~\ref{ape_ssec:modify}, we discuss a modified version of the hydrodynamic equation, but for simplicity, we neglect these modifications here.
Despite this crude approximation, the resulting equation qualitatively explains the phase separation and edge currents, as discussed in the main text. 
Using $g(\bm r,\phi) \simeq \tilde g(r,\theta - \phi)$, Eq.~\eqref{eq:averaged-force1} can be written as
\begin{equation}
g^{(0)}_\alpha(\phi) = -\gamma e_\alpha(\phi),    
\end{equation}
with
\begin{equation}
\gamma = -\int_{0}^{r_\RM{c}}\dd r\int_{-\pi}^{\pi}\dd\theta\ rU'(r) \tilde g(r,\theta)\cos\theta.    
\end{equation}
Likewise, one can calculate $g_{\alpha,\beta}^{(1)}(\phi)$ as
\begin{equation}
g_{\alpha,\beta}^{(1)}(\phi) = B\delta_{\alpha,\beta} + B_2 \cos(2\phi)\sigma_{\alpha,\beta},
\end{equation}
where $\sigma_{x,x} = -\sigma_{y,y}=1$ and $\sigma_{x,y}  = \sigma_{y,x} = 0$, and we define $B$ and $B_2$ as
\begin{align}
& B= -\int_{0}^{r_\RM{c}}\dd r\ r^2U'(r)  g(r), \\
& B_2= - \frac{1}{2}\int_{0}^{r_\RM{c}}\dd r\int_{-\pi}^{\pi}\dd\theta\ r^2U'(r) \tilde g(r,\theta) \cos(2\theta).
\end{align}
Here, $g(r) = \int_{-\pi}^{\pi}\dd\theta\ \tilde g(r,\theta)$.
We further assume that the term involving $B_2$ is negligible, since the $\cos(2\phi)$ term corresponds to a component of the nematic tensor $\mathsf Q$, and terms of order $O(\mathsf Q\nabla\rho)$ are neglected in the final hydrodynamic equations, as discussed in Appendix~\ref{ape_ssec:nematic}.
Under these assumptions, the conditional force can be approximated as
\begin{equation}
\bm G(\bm r,\phi,t)\simeq -\gamma \rho(\bm r,t)\bm e(\phi) - B\nabla\rho(\bm r,t).
\end{equation}
Substituting this expression into Eq.~\eqref{eq:one-body-1} yields Eq.~\eqref{eq:one-body-2} in the main text. 
Note that the conditional force $\bm G$ is related to the Irving--Kirkwood pressure tensor $\mathsf P^\RM{IK}$~\cite{Irving_Kirkwood} through $\int_0^{2\pi}\dd\phi\ \bm G(\bm r,\phi)\Psi(\bm r,\phi) = -\nabla\cdot \mathsf P^\RM{IK}(\bm r)$.
Thus, the approximation for the conditional force can be regarded as a pressure expansion in terms of the density and polarization:
\begin{equation}
\nabla\cdot\mathsf P^\RM{IK}(\bm r) \simeq \gamma \rho(\bm r)\bm p(\bm r) + \frac{1}{\rho\chi}\nabla\rho(\bm r),
\end{equation}
where $\chi = 1/(\rho^2 B)$.
\subsection{Approximation for nematic tensor}
\label{ape_ssec:nematic}
The nematic tensor $\mathsf Q$ is defined by
\begin{equation}
\mathsf  Q(\bm r,t) = \expval{\sum_{j=1}^{N}\qty[\bm e(\phi_j(t))\bm e(\phi_j(t)) - \frac{1}{2}\mathbbm 1 ]\delta(\bm r-\bm r_j(t))}.    
\end{equation}
In the main text, we used Eq.~\eqref{eq:nemtic-1} to express $\mathsf Q$ in terms of the polarization and density field. 
We here derive this approximated formula.

To systematically address the hierarchy in the orientation, we first expand the one-body distribution $\Psi(\bm r,\phi)$ to a Fourier series as~\cite{Bertin2009JPhysA, Speck2020SoftMat}
\begin{align}
&\Psi(\bm r,\phi,t) = \frac{1}{2\pi}\sum_{n\in\mathbb Z}\psi_n(\bm r,t)e^{- i n\phi}, \\
&\psi_n(\bm r,t) = \int_{0}^{2\pi}\dd\phi\ \Psi(\bm r,\phi,t)e^{ i n\phi}.
\end{align}
Each Fourier coefficient then corresponds to each moment of the orientation as
$\psi_0(\bm r) = \rho(\bm r),\ \psi_1(\bm r) = p_x(\bm r) + i p_y(\bm r),\ \psi_2(\bm r) = 2Q_{x,x}(\bm r) + 2i Q_{x,y}(\bm r),\cdots$.
In Fourier space, Eq.~\eqref{eq:one-body-2} reads
\begin{equation}
\begin{split}
\p_t\psi_n = &-\p_z\qty(v(\rho)\psi_{n+1})-\p_{z^*}\qty(v(\rho)\psi_{n-1}) \\ 
&+ \nabla\cdot \psi_n\nabla\psi_0 -D_\RM{r}n^2 \psi_n + in \Omega(\rho)\psi_n \\ 
&- in B \alpha \qty[\psi_{n+1}\p_z + \psi_{n-1}\p_{z^*}]\psi_0
\label{eq:one-body-3}
\end{split}
\end{equation}
with $\p_z = \qty(\p_x - i\p_y)/2$ and $\p_{z^*} = \qty(\p_x + i\p_y)/2$.
Equation~\eqref{eq:one-body-3} with $n=1$ is nothing but Eq.~\eqref{eq:polar-1}.
The solution of Eq.~\eqref{eq:one-body-3} roughly behaves as $\psi_n \sim e^{-n^2D_\RM{r}t + i n \Omega(\rho)t}$, and thus $\psi_n$ decays and oscillates faster for large $n$.
We here assume that $\psi_{\pm n}\simeq 0$ for $n\geq 3$.
In addition, we postulate that the non-linear $\nabla\cdot\psi_{\pm n}\nabla\psi_0$ is negligible for $n\geq 1$.
To obtain an expression for the nematic tensor, we adapt the adiabatic approximation: $\p_t\psi_2(\bm r,t)\simeq 0$.
We then find the quasi-stationary solution
\begin{equation}
\psi_2(\bm r,t) \simeq - \frac{1}{4D_\RM{r} - 2i\Omega(\rho)}\p_{z^*} \qty[v(\rho)\psi_1(\bm r,t)].
\end{equation}
In the tensor representation, this equation means that
\begin{equation}
\mathsf Q(\bm r,t) \simeq -\mathsf A \mathsf q(\bm r,t), \label{eq:nematic_approx}
\end{equation}
where $\mathsf A$ and $\mathsf q(\bm r)$ are defined in Eqs.~\eqref{eq:tensor_q} and \eqref{eq:matrix_A}.
\begin{figure}[t]
\centering
\includegraphics[width=6cm]{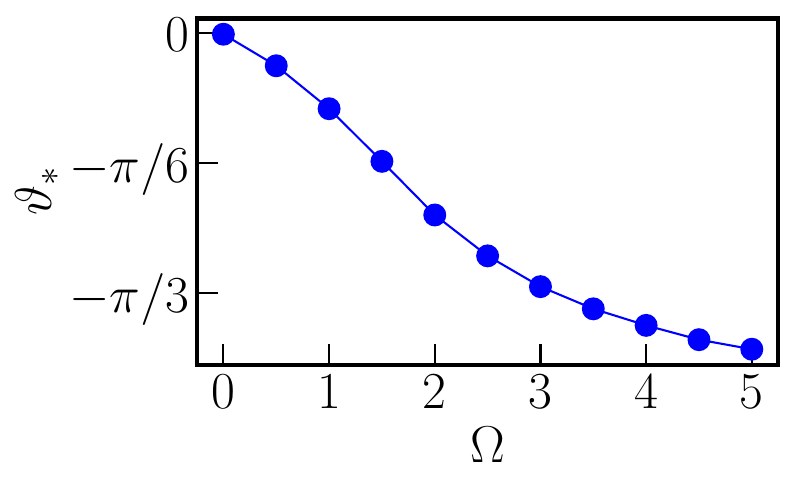}
\caption{\label{fig:theta_star}
$\Omega$ dependence of the angle of the symmetry axis of the pair distribution function with respect to the $x$-axis.
The parameters are set to $\alpha = 0$, $D_{\mathrm{r}} = 0.1$, and $\varphi = 0.4$.
The angle $\vartheta_*$ is extracted by fitting the von Mises distribution
$g(\theta) \propto \exp[k\cos(\theta-\vartheta_*)]$
to $g(r=\sigma,\theta-\phi)$.
}
\end{figure}
\subsection{Modified equations}
\label{ape_ssec:modify}
As discussed in Appendix~\ref{ape_ssec:closure}, the derivation of Eq.~\eqref{eq:one-body-2} is based on a closure approximation.
In this approximation, we used the fact that the pair distribution function is axisymmetric with respect to the orientation $\bm e(\phi)$, at least for small $\Omega$. 
However, this assumption should be modified if $\Omega$ or $\alpha$ is large. 
Here, we briefly discuss how the hydrodynamic equations are modified when the axisymmetry assumption is violated.

In Fig.~\ref{fig:gr}, we show the pair distribution function for several parameter sets.
We measure $g(r,\theta,\phi)$ in a local frame fixed to a tagged particle, as illustrated in Fig.~\ref{fig:gr}(a).
In the achiral case, $\Omega = 0$ and $\alpha=0$, $g(r,\theta,\phi)$ is axisymmetric with respect to the orientation $\bm e(\phi)$ [Figs.~\ref{fig:gr}(b) and (f)], as reported in Refs.~\cite{Bialk2013EPL, Poncet2021PRE}.
However, when $\Omega$ is finite, the symmetry axis is shifted to the right with respect to the vector $\bm e(\phi)$ [Figs.~\ref{fig:gr}(c) and (g)]. 
The shifted angle as a function of $\Omega$ is shown in Fig.~\ref{fig:theta_star}.
Increasing $\alpha$ gives rise to a similar effect, but shifts the symmetry axis to the left with respect to the vector $\bm e(\phi)$ [Figs.~\ref{fig:gr}(e) and (h)]. 
Intuitively, these observations reflect the tendency of chiral particles to push other particles toward the outside of their circular trajectories. 
Taking into account these observations, the assumption $g(\bm r,\phi) = \tilde g(r,\theta - \phi)$ used in Appendix~\ref{ape_ssec:closure} should be modified as follows when $\Omega$ or $\alpha$ is large:
\begin{equation}
g(\bm r,\phi) = \tilde g(r,\theta - \phi - \vartheta_{*}),
\end{equation}
where $\vartheta_{*}$ represents the angle of the axis of symmetry with respect to the $x$-axis, and we also assume that $\tilde g(\cdot, \phi) =\tilde g(\cdot, -\phi)$.
In the case of $\Omega=0$ and $\alpha = 0$, $\vartheta_{*}=0$. 
Equation~\eqref{eq:averaged-force1} then becomes
\begin{equation}
g^{(0)}_\alpha(\phi) = - \gamma_\parallel  e_\alpha(\phi) - \gamma_\perp  e^\perp_\alpha(\phi),
\end{equation}
where $\gamma_\parallel=\gamma\cos\vartheta_{*}$,  $\gamma_\perp=\gamma\sin\vartheta_{*}$, and $\bm e^\perp(\phi) = (\sin\phi,-\cos\phi)$.
Accordingly, the equation for the one-body distribution is modified as
\begin{equation}
\begin{split}
\p_t \Psi = &-\nabla\cdot \Psi\qty[\mathsf v(\rho)\bm e(\phi) - \mu B\nabla\rho] \\
&-\p_\phi\qty[\Omega(\rho) - B\alpha \qty(\bm e(\phi) \cdot \nabla \rho) -D_\RM{r} \p_\phi]\Psi,
\end{split}
\end{equation}
where we introduce
\begin{align}
&v_{\alpha,\beta}(\rho)=v(\rho)\delta_{\alpha,\beta}+v_\perp(\rho)\epsilon_{\alpha,\beta}, \\
&v(\rho) = v_0 - \mu \gamma_\parallel \rho(\bm r), \\
&v_\perp(\rho)=-\mu\gamma_\perp\rho(\bm r), \\
&\Omega(\rho) = \Omega - \alpha \gamma_\parallel \rho(\bm r).
\end{align}
From the same procedure discussed in the main text and above, one can derive the equations for the density field and polarization.
The density field satisfies the continuity equation with the current
\begin{equation}
\bm J(\bm r,t) = \mathsf v(\rho) \bm p(\bm r,t) - D_\RM{e}(\rho) \nabla \rho(\bm r,t).
\end{equation}
In the same way as that shown in Appendix~\ref{ape_ssec:nematic}, one finds the approximated nematic tensor expressed by Eq.~\eqref{eq:nematic_approx} with the modified tensor $\mathsf q(\bm r,t)$ given by
\begin{align}
q_{\alpha,\beta} = \p_\mu v_{\mu,\alpha}(\rho)p_{\beta}+ \p_\mu v_{\mu,\beta}(\rho)p_\alpha - \p_\mu v_{\mu,\nu}(\rho) p_\nu\delta_{\alpha,\beta}.
\end{align}
The final equation for the polarization has the same form as Eq.~\eqref{eq:polar-2}:
\begin{equation}
\begin{split}
\p_t \bm p = &- [D_\RM{r} - \eta (\rho)\nabla^2]\bm p  - \frac{1}{2}\nabla\qty[\rho v(\rho)]  \\ 
&-\qty[ \Omega(\rho) + \eta_\RM{o}(\rho)\nabla^2 ]\bm p^\perp + \frac{1}{2}\nabla^{\perp}\qty[\rho\tilde v_\perp(\rho)],
\end{split}
\end{equation}
where $\tilde v_\perp(\rho) = v_\perp(\rho) + \alpha B \rho/2$, and the viscosity coefficients are modified as
\begin{align}
&\eta(\rho) = A\det[\mathsf v(\rho)] 
= \frac{D_\RM{r}\qty[ v^2(\rho) + v^2_\perp (\rho)]}{16 D_\RM{r}^2 + 4 \Omega^2(\rho)}, \\ 
&\eta_\RM{o} (\rho)  = A_\perp\det[\mathsf v(\rho)]
=\frac{ \Omega(\rho)\qty[ v^2(\rho) + v^2_\perp (\rho)]}{32D_\RM{r}^2 + 8 \Omega^2(\rho)}.
\end{align}
Here, $A$ and $A_\perp$ are the symmetric and antisymmetric parts of Eq.~\eqref{eq:matrix_A}, respectively.
Note that the ratio of the viscosities, $\eta_\RM{o}(\rho)/\eta(\rho) = \Omega(\rho)/(2D_\RM{r})$, is altered by this modification.

As shown in Fig.~\ref{fig:theta_star}, the angular shift is not negligible when $\Omega$ is large, even for $\alpha=0$.
Therefore, the modification introduced here appears to be relevant for large $\Omega$ or $\alpha$.
However, as discussed in the main text, the occurrence of phase separation and the presence of edge currents along the interface can be qualitatively explained without incorporating $\vartheta_*$.
We therefore infer that this modification is not crucial, at least for the qualitative mechanism of chirality-induced phase separation and edge currents.
In a recent work~\cite{Metzger2026}, the perpendicular component of the effective self-propulsion was also taken into account for a quantitative prediction of the edge flux in a different model of chiral particles.
We leave a quantitative assessment for future work.

\subsection{Scalar field theory for odd fluids}
\label{ape_ssec:field_theory}
Here, we briefly discuss a possible connection between our hydrodynamic equations and a scalar field theory for odd fluids.
For active fluids, scalar field theories, such as active Model B(+), have often been used to study large-scale collective behavior~\cite{Wittkowski2014NatCom, Tjhung2018PRX, Saha2020PRX, Speck2022PRE}.
Even for chiral active fluids, several studies have proposed descriptions based on a scalar variable~\cite{Bickmann2022JCP, Kalz2024JPhysA, wang2026edge}.
In the present system, since the polarization is a fast variable, the behavior at the largest scales should also be describable solely in terms of a conserved scalar variable.
Toward such a field theory, we first rewrite the first term of Eq.~\eqref{eq:current-many1} as~\cite{Speck2015JCP}
\begin{equation}
\mathcal D(\rho)\nabla\rho(\bm r) = \nabla\fdv{\mathcal F_\RM{bulk}[\phi]}{\phi(\bm r)},   
\end{equation}
where we introduce $\phi(\bm r) = \rho(\bm r) - \rho_0$ and the bulk ``free-energy'' functional
\begin{equation}
\mathcal F_\RM{bulk}[\phi] = \int_V\dd[2]\bm r\qty[\frac{c_2}{2!}\phi^2(\bm r) +\frac{c_3}{3!}\phi^3(\bm r)
+\frac{c_4}{4!}\phi^4(\bm r) + \cdots].    
\end{equation}
Each coefficient $c_k$ is determined from the density expansion of $\mathcal D(\rho)$ around $\rho_0$; for example, $c_2 = \mathcal D(\rho_0)$.
In deriving the hydrodynamic equations in Sec.~\ref{sec:collective_behavior} and Appendix~\ref{ape_ssec:closure}, we truncated the gradient expansion at $O(\nabla)$.
However, including higher-order contributions of $O(\nabla^3)$ can generate the stiffness term
\begin{equation}
\mathcal F_\RM{grad}[\phi] = \int_V\dd[2]\bm r\ \frac{\kappa}{2}|\nabla\phi(\bm r)|^2,
\end{equation}
and the ``odd stiffness'' term $\kappa_\RM{o}(\phi)\nabla^\perp\nabla^2\phi$.
Here, $\kappa$ is assumed to be constant for simplicity.
Therefore, the simplest form of the equation up to $O(\nabla^3)$ can be written as $\p_t\phi = -\nabla\cdot \bm J$ with
\begin{equation}
\bm J = -\nabla\fdv{\mathcal F[\phi]}{\phi} - \mathcal D_\RM{o}(\phi)\nabla^\perp\phi +\kappa_\RM{o}(\phi)\nabla^\perp\nabla^2\phi,  
\end{equation}
where $\mathcal F[\phi] =\mathcal F_\RM{bulk}[\phi]+ \mathcal F_\RM{grad}[\phi]$.
As discussed in Sec.~\ref{sec:collective_behavior}, the term involving the odd diffusivity $\mathcal D_\RM{o}(\phi)$ can generate edge currents when the system possesses interfaces, but does not affect the profile of $\phi$ owing to its divergence-free nature.
Indeed, substituting the current $\bm J$ into the continuity equation yields
\begin{equation}
\p_t\phi = \nabla^2\fdv{\mathcal F[\phi]}{\phi} + \kappa'_\RM{o}(\phi)\qty{\nabla^2\phi,\phi }, \label{eq:odd_ModelB}
\end{equation}
where $\qty{A,B}= (\p_xA) (\p_yB) - (\p_yA) (\p_xB)$ and the prime in $\kappa'_\RM{o}(\phi)$ denotes differentiation with respect to $\phi$.
The odd stiffness term can affect the density profile when it depends on $\phi$, and thus may alter the interface shape, although this effect appears to be small in our system (see the density profile shown in Fig.~\ref{fig:collective_current}).
We also remark that the second term in Eq.~\eqref{eq:odd_ModelB} is nonlinear; therefore, the linearized equation contains no odd contribution.
This means that the edge currents cannot be discussed from the perspective of topology based on the linearized equation, unlike in Sec.~\ref{sec:topo}.
Studying the topological properties of scalar field theories with nonlinear odd terms would be an important direction for future work.

\begin{figure}[b!]
\centering
\includegraphics[width=1.0\linewidth]{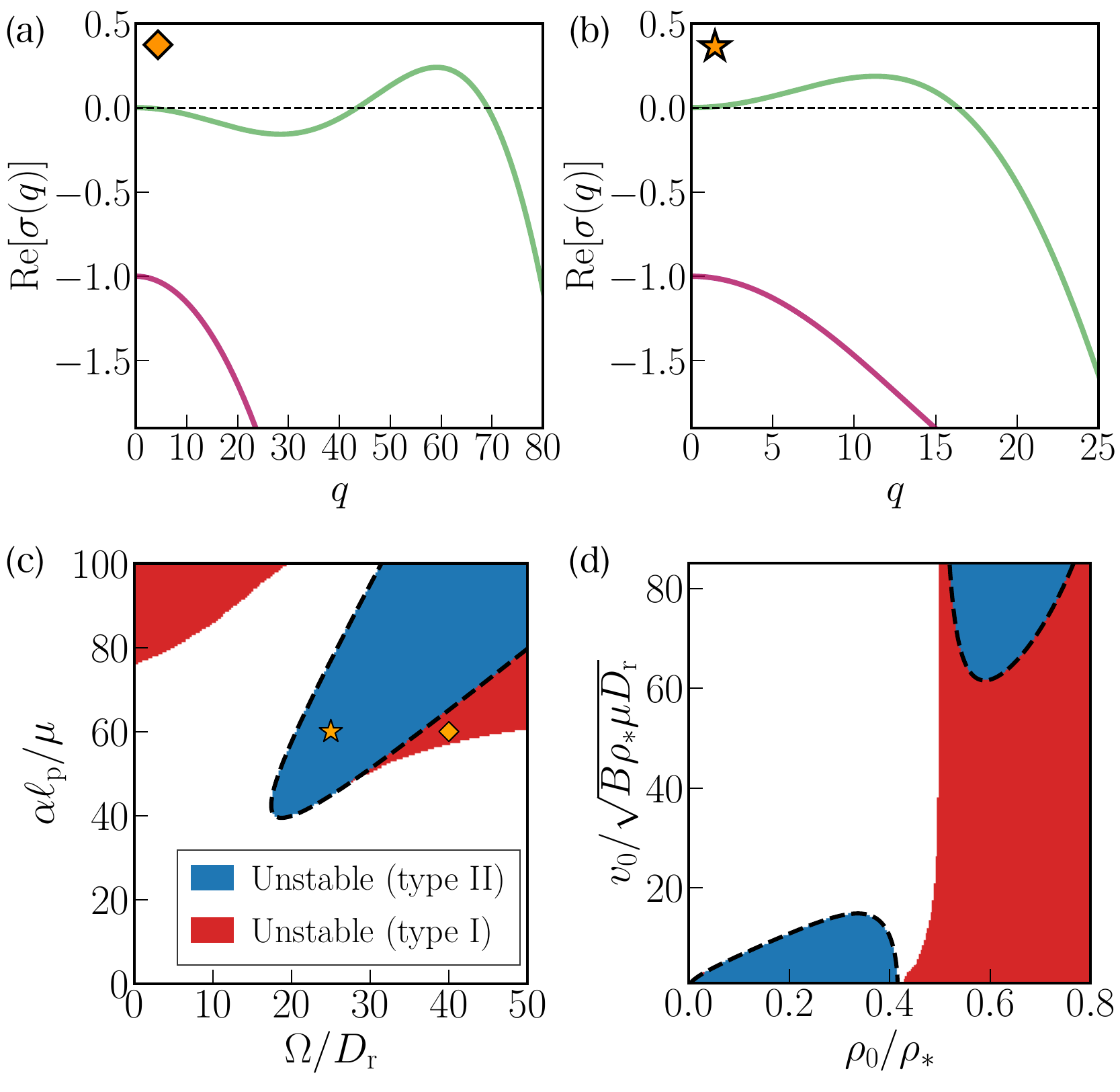}
\caption{\label{fig:eigenvalues}
Real part of the dispersion relation at (a)~$\Omega/D_\RM{r}=40$ and (b)~$\Omega/D_\RM{r}=25$.
The other parameters are set to
$\alpha\ell_\RM{p}/\mu =60$, $v_0/\sqrt{B\rho_{*}\mu D_\RM{r}} = 10$, and $\rho_0/\rho_* = 0.3$.
(c)~Linearly unstable region at fixed $v_0/\sqrt{B\rho_{*}\mu D_\RM{r}} = 10$ and $\rho_0/\rho_* = 0.3$.
(d)~Linearly unstable region at fixed $\alpha\ell_\RM{p}/\mu =60$ and $\Omega/D_\RM{r}=25$.
The regions in which the instability sets in at infinitesimally small $q$ (type~II) and at finite $q$ (type~I) are shown in blue and red, respectively.
The white region represents the linearly stable regime.
The black dashed lines represent the solutions of $\mathcal D(\rho_0) = 0$ [see Eq.~\eqref{eq:collective_diffusion} in the main text].
The diamond and star markers in panel~(c) indicate the parameter values corresponding to panels~(a) and (b), respectively.
}
\end{figure}
\section{Linear analysis of the hydrodynamic equations}
\label{ape:linear}
\subsection{Linear stability}
\label{ape_ssec:stability}
In Sec.~\ref{sec:collective_behavior}, we considered the linear stability of our hydrodynamic equations at the level of the adiabatic approximation.
Here, we briefly discuss the linear stability of the coupled equations for the density and polarization.
For simplicity, we work with a dimensionless set of equations.
Using the length and time scales $\ell_\RM{p} = v_0/D_\RM{r}$ and $1/D_\RM{r}$, respectively, we rescale position and time as $\bm r\to \bm r/\ell_\RM{p}$ and $t\to t D_\RM{r}$.
We also define the reference density $\rho_{*} ={v_0}/(\mu\gamma)$ and rescale the density and polarization as $\rho\to \rho/\rho_*$ and $\bm p\to \bm p/\rho_*$, respectively.
After nondimensionalization, we linearize the continuity equation with Eq.~\eqref{eq:density_current_1} and Eq.~\eqref{eq:polar-2} around the homogeneous solution, $\rho(\bm r,t) = \rho_0 + \delta\rho(\bm r,t)$ and $\bm p(\bm r,t) = \bm 0 + \delta\bm p(\bm r,t)$.
We then Fourier-transform the equations and apply the Helmholtz decomposition to the polarization $\delta\hat {\bm p}(\bm q)$ as
\begin{equation}
\delta\hat {\bm p}(\bm q) = \delta\hat p_\RM{L}(\bm q)\bm e_\RM{L} + \delta\hat p_\RM{T}(\bm q)\bm e_\RM{T},
\end{equation}
where $\bm e_\RM{L}$ and $\bm e_\RM{T}$ denote the unit vectors parallel and perpendicular to the wavevector $\bm q$, respectively.
Since the system is rotationally invariant, one can take $\bm q = q\bm e_x = q\bm e_\RM{L}$ without loss of generality.
The equation for the set of field variables, $\bm \psi :=(\delta\rho,\delta p_\RM{L}, \delta p_\RM{T})$, then reads
\begin{align}
\p_t  \hat{\bm \psi}(\bm q,t ) =  {\mathsf G}(q)\hat{\bm \psi}(\bm q,t ) \label{eq:linear-1}
\end{align}
with
\begin{align}
\mathsf G(q)
= 
\mqty( - b q^2 & -i\lambda q & 0 \\ 
-i\lambda_2 q & -(1+\nu q^2) & -(\mathcal W - \nu_{\rm o}q^2) \\ 
-i\lambda_\perp q & (\mathcal W - \nu_{\rm o}q^2) & -(1+\nu q^2) ).
\end{align}
Here, the dimensionless parameters are defined by $\lambda= (1-\rho_0/\rho_*)$, $\lambda_2 = (1-2\rho_0/\rho_*)/2$, 
$b = {B \rho_0 \mu  D_\RM{r}}/{v_0^2}$, 
$\lambda_\perp = \alpha b\ell_\RM{p}\rho_0/(2\mu \rho_*)$, 
$\mathcal W = \Omega/D_\RM{r} - \alpha\ell_\RM{p}\rho_0/(\mu\rho_*)$, 
$\nu = \eta(\rho_0)/(\ell_\RM{p}^2D_\RM{r})$, 
and 
$\nu_\RM{o} = \eta_\RM{o}(\rho_0)/(\ell_\RM{p}^2D_\RM{r})$.
The eigenvalues of the matrix $\mathsf G(q)$, denoted by $\sigma(q)$, give the dispersion relation.
In particular, the real part of $\sigma(q)$ characterizes the growth rate of small fluctuations around the homogeneous solution.
If there is at least one eigenvalue satisfying $\RM{Re}[\sigma(q)]>0$, the homogeneous solution is linearly unstable.

Figures~\ref{fig:eigenvalues}(a) and (b) show two different types of instability.
In Fig.~\ref{fig:eigenvalues}(a), the instability sets in at a finite $q$, whereas in Fig.~\ref{fig:eigenvalues}(b), it sets in at infinitesimally small $q$.
These two instabilities are referred to as ``type~I'' and ``type~II,'' respectively~\cite{Cross1993RevModPhys}, and the type~II instability corresponds to macrophase separation.
In Figs.~\ref{fig:eigenvalues}(c) and (d), we show the regions of type~I and type~II instability in the $\Omega$-$\alpha$ and $\rho_0$-$v_0$ planes, respectively.
The black dashed lines represent the solutions of $\mathcal D(\rho_0) = 0$, which are the same as those shown in Fig.~\ref{fig:instability} of the main text.
The type~I instability, shown by the red regions in Figs.~\ref{fig:eigenvalues}(c) and (d), cannot be captured by the adiabatic approximation and may correspond to states in which multiple small clusters are present.
Clustering states corresponding to a type~I instability in chiral active Brownian particles have also been reported in a previous study~\cite{Ma2022JCP}.

\subsection{Effective Hamiltonian}
\label{ape_ssec:Hamiltonian}
Equation~\eqref{eq:linear-1} (without the Helmholtz decomposition) can be rewritten as
$i\p_t\hat {\bm \psi}(\bm q,t)  = \mathsf H(\bm q)\hat {\bm \psi}(\bm q,t)$
with the Hamiltonian-like matrix $\mathsf H(\bm q) = i\mathsf G(\bm q)$:
\begin{align}
\mathsf H(\bm q) =
\mqty( -i b q^2 & \lambda q_x & \lambda q_y \\
\lambda_2 q_x-\lambda_\perp q_y & -i(1+\nu q^2) & -i(\mathcal W-\nu_{\rm o}q^2) \\ 
\lambda_2 q_y+\lambda_\perp q_x & i(\mathcal W-\nu_{\rm o}q^2) & -i(1+\nu q^2) ). \label{eq:Hamiltonian1}
\end{align} 
We now rescale the field variable as 
$\hat{\bm \psi}(\bm q) = \mathsf S \hat{\bm \psi}'(\bm q)$ with $\mathsf S = \RM{diag}(1,s,s)$.
The equation is then transformed as $i\p_t  \hat{\bm \psi}'(\bm q,t ) = \tilde {\mathsf H}(\bm q)\hat{\bm \psi}'(\bm q,t ) ,\ \tilde {\mathsf H}(\bm q)= \mathsf S^{-1}\mathsf H\mathsf S$.
If we choose $s = \sqrt{\lambda_2/\lambda}$, the rescaled matrix reads
\begin{align}
\tilde{\mathsf H}(\bm q) = \mqty( -i b q^2 & v_\RM{s} q_x & v_\RM{s} q_y \\ 
v_\RM{s} q_x - \Lambda_\perp q_y & -i(1+\nu q^2) & -i(\mathcal W - \nu_{\rm o}q^2) \\ 
v_\RM{s} q_y + \Lambda_\perp q_x & i(\mathcal W - \nu_{\rm o}q^2) & -i(1+\nu q^2) ),
\end{align}
where $v_\RM{s}= \sqrt{\lambda \lambda_2}$ and $ \Lambda_\perp = \lambda_\perp\sqrt{{\lambda}/{\lambda_2}}$. 
The Hermitian part of $\tilde{\mathsf H}(\bm q)$ is given by
\begin{align}
\mathsf H_0(\bm q) = d_x(\bm q)\mathsf L_x + d_y(\bm q)\mathsf L_y + d_z(\bm q)\mathsf L_z =:\bm d(\bm q)\cdot{\bf L} \label{eq:Hamiltonian2}
\end{align}
where $\bm d(\bm q) =  (v_\RM{s}q_x, v_\RM{s}q_y,  \mathcal W-\nu_{\rm o}q^2)$, ${\bf L} = (\mathsf L_x, \mathsf L_y, \mathsf L_z)$, and 
\begin{align}
\mathsf L_x =\mqty(0 & 1 & 0 \\ 1 & 0 & 0 \\ 0 & 0&0 ),\ 
\mathsf L_y =\mqty(0 & 0 & 1 \\ 0 & 0 & 0 \\ 1 & 0& 0 ),\ 
\mathsf L_z =\mqty(0 & 0 & 0 \\ 0 & 0 & -i \\ 0 & i &0 ).
\end{align}
This matrix $\mathsf H_0(\bm q)$ has been used to discuss topological edge modes in odd fluids~\cite{Souslov2019PRL, Fujii2025SciPost}.

\subsection{Chern numbers}
\label{ape_ssec:Chern}
In Refs.~\cite{Souslov2019PRL, Fujii2025SciPost}, the Chern numbers were calculated for systems described by the Hermitian matrix given by Eq.~\eqref{eq:Hamiltonian2}.
In this section, we calculate the Chern numbers from the effective Hamiltonian, Eq.~\eqref{eq:Hamiltonian1}, and show that they are identical to those calculated from the Hermitian part in Eq.~\eqref{eq:Hamiltonian2}.

Our effective Hamiltonian is non-Hermitian, i.e., $\tilde{\mathsf H}(\bm q) \neq \tilde{\mathsf H}^\dagger(\bm q)$.
Even for non-Hermitian systems, Chern numbers can be defined for separable bands~\cite{Shen2018PRL}.
Let $\ket{u_n^\RM{R}(\bm q)}$ and $\ket{u_n^\RM{L}(\bm q)}$ be the $n$-th right and left eigenvectors, respectively, satisfying
\begin{align}
&\tilde{\mathsf H}(\bm q) \ket{u_n^\RM{R}(\bm q)} = E_n(\bm q)\ket{u_n^\RM{R}(\bm q)}, \\ 
&\tilde{\mathsf H}^\dagger(\bm q) \ket{u_n^\RM{L}(\bm q)} = E_n^{*}(\bm q)\ket{u_n^\RM{L}(\bm q)}.
\end{align}
We impose the separability condition on the $n$-th eigenvalue: $E_n(\bm q) \neq E_m(\bm q)$ for all $m(\neq n)$ and all $\bm q$.
The Berry connection can then be defined as
\begin{equation}
\bm A_n^\RM{LR}(\bm q) = i\bra{u_n^\RM{L}(\bm q)}\p_{\bm q}\ket{u_n^\RM{R}(\bm q)}.
\end{equation}
The Chern number is then given by
\begin{equation}
\mathcal C_n^\RM{LR} = \frac{1}{2\pi}\int\dd[2]\bm q\ F_n^\RM{LR}(\bm q),
\end{equation}
where
\begin{equation}
F_n^\RM{LR}(\bm q) = \qty[\p_{\bm q}\times \bm A_n^\RM{LR}(\bm q) ]\cdot \bm e_z
\end{equation}
is the Berry curvature.
One can also define $\mathcal C_n^\RM{LL}$, $\mathcal C_n^\RM{RR}$, and $\mathcal C_n^\RM{RL}$, but it has been proven that all these Chern numbers take the same value~\cite{Shen2018PRL}.
Therefore, we henceforth consider only $\mathcal C_n^\RM{LR}$ and omit the index ``$\RM{LR}$'' for simplicity of notation.

To calculate the Chern numbers for our system, it is convenient to use the fact that the angular dependence in Eq.~\eqref{eq:Hamiltonian1} can be eliminated by a unitary transformation.
We denote the wavevector $\bm q$ in polar coordinates as $\bm q  = q(\cos\phi,\sin\phi)$.
Using the unitary matrix
\begin{equation}
\mathsf U_\phi = e^{-i\phi \mathsf L_z},
\end{equation}
Eq.~\eqref{eq:Hamiltonian1} can be transformed as
\begin{equation}
\mathsf U_\phi ^\dagger \tilde{\mathsf H}(q, \phi)\mathsf U_\phi = \tilde{\mathsf H}(q,0) =: \tilde{\mathsf H}(q). \label{eq:unitary1}
\end{equation}
In addition, the right and left eigenvectors satisfy
\begin{align}
&\ket{u_n^\RM{R}(q,\phi)} = \mathsf U_\phi\ket{u_n^\RM{R}(q)}, \label{eq:unitary2}\\
&\bra{u_n^\RM{L}(q,\phi)} = \bra{u_n^\RM{L}(q)}\mathsf U_\phi^\dagger, \label{eq:unitary3}
\end{align}
where $\ket{u_n^\RM{R}(q)}$ and  $\ket{u_n^\RM{L}(q)}$ are right and left eigenvectors of $\tilde{\mathsf H}(q)$, respectively. 
The Berry connection in the polar coordinates is expressed as 
\begin{equation}
\bm {A}_{n}(\bm q) 
= \bm e_{q} A^{(q)}_{n}(\bm q) + \frac{1}{q}\bm e_{\phi} A^{(\phi)}_{n}(\bm q),
\end{equation}
where $A^{(q)}_{n}(\bm q) =  i \bra{u_{n}^\RM{L}(\bm q)}\p_{q}\ket{u_{n}^\RM{R}(\bm q)}$,\ $
A^{(\phi)}_{n}(\bm q) =  i \bra{u_{n}^\RM{L}(\bm q)}\p_{\phi}\ket{u_{n}^\RM{R}(\bm q)}$.
From Eqs.~\eqref{eq:unitary2} and \eqref{eq:unitary3}, one can show that $\bm {A}_{n}(\bm q)$ is independent of $\phi$, and
\begin{equation}
A^{(\phi)}_{n}(q)  = \bra{u_{n}^\RM{L}(q)}\mathsf L_z\ket{u_{n}^\RM{R}(q)}. \label{eq:Beery_conne1}
\end{equation}
The Berry curvature in polar coordinates is therefore given by $F_{n}(q)  = [\p_q A^{(\phi)}_{n}(q)]/q$.
Thus, the Chern numbers are calculated as
\begin{align}
\mathcal C_{n} &=
 \int_{0}^{\infty}\dd q\ \p_q A^{(\phi)}_{n}(q) \notag \\
& = \lim_{q\to\infty}A^{(\phi)}_{n}(q) - \lim_{q\to 0}A^{(\phi)}_{n}(q).
\end{align} 
This formula implies that, for systems satisfying Eq.~\eqref{eq:unitary3}, it is sufficient to evaluate eigenvectors in the small- and large-$q$ limits to obtain $\mathcal C_{n}$.

In the limit $q\to 0$, the eigenvalues of Eq.~\eqref{eq:Hamiltonian1} are given by
$E_0(q\to 0) = 0$ and $E_\pm(q\to 0) = -i \pm |\mathcal W|$.
For large $q$, $E_0(q\gg 1) = - ib q^2$ and $E_\pm(q\gg 1) = - i\nu q^2 \pm |\nu_\RM{o}|q^2$.
The eigenvectors corresponding to $E_\pm(q)$ in these two limits are obtained as
\begin{align}
\ket{u_\pm^\RM{R}(q)} &= \ket{u_\pm^\RM{L}(q)} \notag \\
&=
\begin{dcases}
\frac{1}{\sqrt{2}}\mqty[0 & 1 & \pm i\RM{sgn}(\mathcal W)]^\top & (q \to 0), \\
\frac{1}{\sqrt{2}}\mqty[0 & 1 & \mp i\RM{sgn}(\nu_\RM{o})]^\top & (q \to \infty).
\end{dcases}
\end{align}
Note that the right and left eigenvectors coincide in the small- and large-$q$ limits, i.e., the non-Hermiticity is irrelevant in these limits.
The Berry connection is obtained by using  Eq.~\eqref{eq:Beery_conne1} as
\begin{align}
A_{\pm}^{(\phi)}(q)=
\begin{dcases}
\pm\RM{sgn}(\mathcal W) & (q \to 0), \\
\mp\RM{sgn}(\nu_\RM{o}) & (q \to \infty),
\end{dcases}
\end{align}
leading to the Chern numbers:
\begin{align}
\mathcal C_{\pm} = \mp\qty[ \RM{sgn}(\nu_\RM{o}) + \RM{sgn}(\mathcal W)].
\end{align} 
This is equivalent to that obtained in Refs.~\cite{Souslov2019PRL, Fujii2025SciPost}.
Returning to the original dimensions and using the microscopic expression for the odd viscosity, Eq.~\eqref{eq:odd_viscous1}, we obtain
\begin{align}
\mathcal C_{\pm} &= \mp\qty{ \RM{sgn}[\eta_\RM{o}(\rho_0)] + \RM{sgn}[\Omega(\rho_0)]} \notag \\
&= \mp 2\RM{sgn}[\Omega(\rho_0)]. 
\end{align} 

\begin{figure*}[t]
\centering
\includegraphics[width=1.0\linewidth]{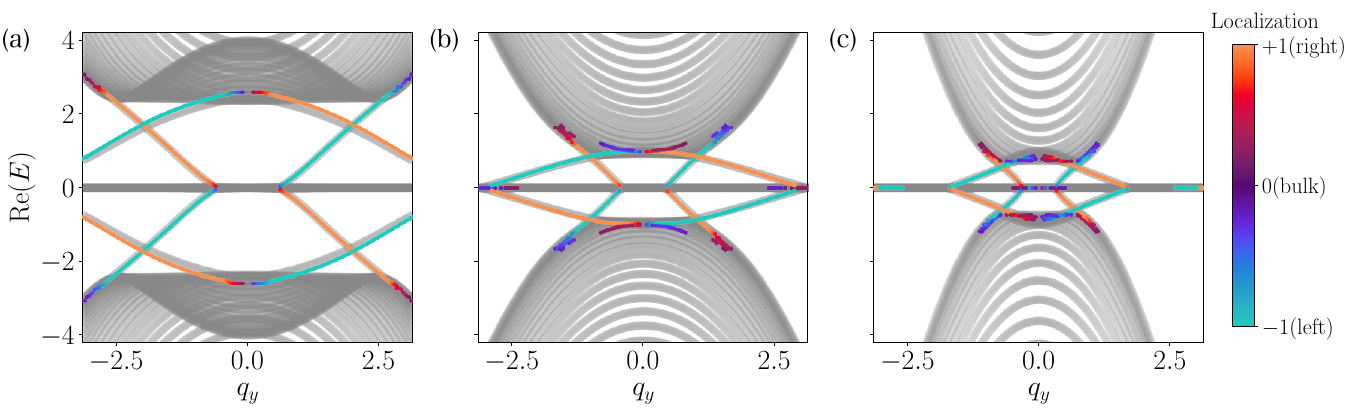}
\caption{\label{fig:edge_solutions} Edge modes and band structure of the effective Hamiltonian. The band structure (gray lines) is obtained in the same manner as in Fig.~\ref{fig:band}, using systems with periodic boundary conditions in the $y$-direction and open boundary conditions in the $x$-direction. The edge modes are superimposed on the bands as colored lines. These edge modes are obtained independently from a physically motivated ansatz that captures their decay from the walls into the bulk of the system (see Appendix~\ref{ape_ssec:EdgeModes} for details). From left to right, the parameters are set to (a)~$b=0.1$, $v_\RM{s} = 1$, $\Lambda_\perp = 0.1$, $\nu = 0.25$, $\nu_\RM{o}=0.5$, and $\mathcal W=4$, (b)~$b=0.1$, $v_\RM{s} = 1$, $\Lambda_\perp = 0.1$, $\nu = 0.2$, $\nu_\RM{o}=0.5$, $\mathcal W=1$, and (c)~$b=0.2$, $v_\RM{s} = 1$, $\Lambda_\perp = 0.3$, $\nu = 0$, $\nu_\RM{o}=1$, $\mathcal W=1$. In all plots, the $x$-direction is discretized into $M=100$ sites with $\dd x=0.2$, and a domain width of $L=20$ is used in the edge-mode calculation.
}
\end{figure*}

\subsection{Band structure}
\label{ape_ssec:BandStructure}

Here, we provide the technical details of the numerical calculation of the band structure shown in Fig.~\ref{fig:band} in the main text.
To obtain the band structure, we consider a system that is periodic in the $y$-direction and open in the $x$-direction, following Ref.~\cite{Sone2020NatComm}.
The effective Hamiltonian in real space reads
\begin{equation}
\mathcal H
=
\mqty( i b\nabla^2 & -i v_\RM{s} \p_x & -i v_\RM{s} \p_y 
\\ -iv_\RM{s} \p_x+i\Lambda_\perp \p_y & -i(1 - \nu \nabla^2) & -i(\mathcal W + \nu_{\rm o}\nabla^2) \\ -iv_\RM{s} \p_y-i\Lambda_\perp \p_x & i(\mathcal W + \nu_{\rm o}\nabla^2) & -i(1 - \nu \nabla^2) ). \label{eq:hamiltonian_real}
\end{equation}
Since the system is periodic in the $y$-direction, we replace the spatial derivatives $\partial_y$ and $\nabla^2$ with $iq_y$ and $\partial_x^2-q_y^2$, respectively.
We then discretize the $x$-direction into $M$ sites with lattice spacing $\dd x$ and approximate $\partial_x$ and $\partial_x^2$ using the standard central-difference scheme.
Finally, we numerically compute the eigenvalues of the resulting $3M\times 3M$ matrix.

To quantify the degree to which each eigenmode is localized near an edge, we also calculate the degree of localization of the eigenvectors.
Consider a specific mode $n\in\{1,2,...,3M\}$ whose normalized eigenvector is given by $u_n = (\{\rho_n(x)\}_{x=1}^{M},\{p^{(x)}_n(x)\}_{x=1}^{M},\{p^{(y)}_n(x)\}_{x=1}^{M})$.
Note that we consider only right eigenvectors here.
We denote elements of the $n$-th eigenvector corresponding to the discretized position $x\in\{1,2,...,M\}$ by $\psi_n(x) = (\rho_n(x),p^{(x)}_n(x), p^{(y)}_n(x))$.
In analogy with quantum mechanics, we interpret $\abs{\psi_n(x)}^2$ as the probability density of finding mode $n$ at site $x$.
We define the $n_\RM{e}$ sites nearest to each boundary as the edge regions of the system.
The localizations at the left and right edges are then defined as
$P^\RM{L}_n = \sum_{x=1}^{n_\RM{e}}\abs{\psi_n(x)}^2$
and
$P^\RM{R}_n = \sum_{x=M-n_\RM{e}+1}^{M}\abs{\psi_n(x)}^2$,
respectively.
We quantify the edge localization by the difference $P^\RM{R}_n-P^\RM{L}_n$.
For modes completely localized at the right or left edge, this quantity takes the value $+1$ or $-1$, respectively, and it is close to zero for bulk modes.
We note that $P^\RM{R}_n-P^\RM{L}_n$ may also be zero for modes having comparable weights at the two edges.
In Fig.~\ref{fig:band} of the main text, we set $\dd x=0.2$, $M=100$, and $n_\RM{e}=4$.

\subsection{Edge modes}
\label{ape_ssec:EdgeModes}

In the main text, we obtain the topological edge modes by discretizing the matrix $\mathcal{H}$ under half-periodic boundary conditions.
Here, we show that the edge modes can also be obtained semi-analytically by starting from an ansatz for the edge solutions and imposing appropriate boundary conditions.
We consider the eigenvalue problem $\mathcal H \bm{\varphi}(x,y) = E\bm{\varphi}(x,y)$ for a fluid confined between rigid walls at $x=-L$ and $x=+L$.
Here, $\mathcal H$ is given by Eq.~\eqref{eq:hamiltonian_real}.
For a physical solution to exist, we expect the edge modes to decay as one moves from either the left or the right wall into the bulk of the system.
Hence, we assume the following form for the edge modes:
\begin{align}
\bm{\varphi}(x,y) = \bm\Phi e^{\kappa  x}e^{iq_y y}.
\end{align}
Substituting this ansatz into the equation $\mathcal H \bm{\varphi}(x,y) = E\bm{\varphi}(x,y)$, we obtain
\begin{align}
\tilde{\mathsf H}(i\kappa, q_y)\bm\Phi  = E \bm\Phi,
\end{align}
where $\tilde{\mathsf H}(i\kappa, q_y)$ is the matrix in Eq.~\eqref{eq:Hamiltonian} with $q_x = i\kappa$.
Hereafter, we simply denote this matrix by $\tilde{\mathsf H}(\kappa)$.
The characteristic polynomial of $\tilde{\mathsf H}(\kappa)$ yields
\begin{align} 
    \tilde{E}(\Gamma^2 - w^2) + v_\RM{s} K^2 (v_\RM{s} \Gamma - i\Lambda_\perp w) = 0,
\end{align}
where 
$K^2 = \kappa^2 - q_y^2$, $\tilde{E} = E - ibK^2$, $\Gamma = E + i(1 - \nu K^2)$ and $w = \mathcal{W} + \nu_o K^2$.
Since this is a cubic equation in $\kappa^2$, we obtain six solutions, $\kappa = \pm\kappa_1,\pm\kappa_2,\pm\kappa_3$.
For any root $\kappa_j$, the corresponding eigenvector can be written as follows after some algebra:
\begin{align}
    \bm\Phi_j = 
    \mqty(  
    v_\RM{s}(q_y \Gamma_j - \kappa_j w_j) \\ 
    -iw_j\tilde{E}_j - v_\RM{s} q_y(iv_\RM{s}\kappa_j + \Lambda_\perp q_y) \\ 
    v_\RM{s}^2\kappa_j^2 - iv_\RM{s}\Lambda_\perp \kappa_j q_y + \Gamma_j\tilde{E}_j 
    ).
\end{align}
The general solution can then be written as
\begin{align}
\bm{\psi}(x,y) = \sum_{j=1}^{6} C_j \bm \Phi_j e^{\kappa_j x} e^{iq_y y} =: \bm{\psi}(x)e^{iq_y y}
\end{align}
where $\{C_j\}$ are determined by the boundary conditions. 
Note that, in the summation, we use $\kappa_j$ to denote the elements of $\{\pm\kappa_1,\pm\kappa_2,\pm\kappa_3\}$.
The modes localized at the left (right) wall $x=-L$ ($x=+L$) should satisfy $\RM{Re}(\kappa)<0$ ($\RM{Re}(\kappa)>0$) so that they decay into the bulk.
We take $\RM{Re}(+\kappa_\alpha)<0$ $(\alpha=1,2,3)$, corresponding to the left edge modes, and define the eigenvectors through $\tilde{\mathsf H}(+\kappa_\alpha)\bm\Phi_\alpha^\RM{L}  = E_\alpha \bm\Phi^\RM{L}_\alpha$.
Likewise, we define the right edge modes through $\tilde{\mathsf H}(-\kappa_\alpha)\bm\Phi_\alpha^\RM{R}  = E_\alpha \bm\Phi^\RM{R}_\alpha$.
Then, $\bm{\psi}(x)$ can be decomposed as
\begin{align}
\bm{\psi}(x) = \sum_{\alpha=1}^{3} A_\alpha \bm\Phi_\alpha^\RM{L}  e^{\kappa_\alpha x} + \sum_{\alpha=1}^{3} B_\alpha \bm\Phi_\alpha^\RM{R}  e^{-\kappa_\alpha x}
\end{align}
As boundary conditions, we impose $\bm{\psi}(-L) = \bm{\psi}(+L) = \bm 0$.
We also rescale coefficients as $A'_\alpha = A_\alpha e^{-\kappa_\alpha L}$ and $B'_\alpha = B_\alpha e^{-\kappa_\alpha L}$.
We introduce this rescaling to improve the numerical stability of the computation described below.
Applying the boundary conditions then yields six equations, which can be written in the following matrix form:
\begin{align}
\mathsf{M}_\RM{BC} \mqty( \bm{A}' \\ \bm{B}')
:= 
\mqty(\mathsf M_\RM{L}^{-} & \mathsf M_\RM{R}^{-} \\
\mathsf M_\RM{L}^{+} & \mathsf M_\RM{R}^{+}
)\mqty( \bm{A}' \\ \bm{B}') = \mathsf O,
\end{align} 
where $\bm{A}' = (A'_1, A'_2, A'_3)^\top$ and $\bm{B}' = (B'_1, B'_2, B'_3)^\top$ and
\begin{align}
&\mathsf M_\RM{L}^{-} = \mqty(\bm\Phi_1^\RM{L} & \bm\Phi_2^\RM{L} & \bm\Phi_3^\RM{L}) \\
&\mathsf M_\RM{R}^{-} = \mqty(\bm\Phi_1^\RM{R}e^{2\kappa_1L} & \bm\Phi_2^\RM{R}e^{2\kappa_2L} & \bm\Phi_3^\RM{R}e^{2\kappa_3L})\\
&\mathsf M_\RM{L}^{+} = \mqty(\bm\Phi_1^\RM{L}e^{2\kappa_1L} & \bm\Phi_2^\RM{L}e^{2\kappa_2L} & \bm\Phi_3^\RM{L}e^{2\kappa_3L}) \\
&\mathsf M_\RM{R}^{+}  = \mqty(\bm\Phi_1^\RM{R} & \bm\Phi_2^\RM{R} & \bm\Phi_3^\RM{R}).
\end{align}
The matrices $\mathsf M^{\pm}_\RM{L/R}$ correspond to the boundary conditions at $x=\pm L$.

The dispersion relation of edge modes $E=E(q_y)$ are obtained by solving $\det[\mathsf{M}_\RM{BC}(E, q_y)] = 0$. 
To obtain the edge modes numerically, we first guess $E$ for each $q_y$ and then calculate $\kappa_j(q_y,E)$.
Using the resulting $\{\kappa_j\}$, we construct the matrix $\mathsf{M}_\RM{BC}$ and check whether it satisfies $\det[\mathsf{M}_\RM{BC}]=0$.
After finding $E(q_y)$, we repeat the same procedure for $q_y+\dd q_y$, using the value of $E$ obtained at $q_y$ as the initial guess.
Figure~\ref{fig:edge_solutions} shows the edge solutions obtained using the procedure described above.
For comparison, the eigenmodes obtained from the direct numerical calculation described in Appendix~\ref{ape_ssec:BandStructure} are also shown as gray lines.
The edge modes obtained by the two different methods are in quantitative agreement.
\bibliography{refs}
\end{document}